\documentclass{snompy}

\usepackage[backend=biber,style=nature]{biblatex}
\usepackage{float}
\usepackage{tabularx}
\usepackage{cprotect}

\usepackage{graphicx}
\graphicspath{{figures/}}

\usepackage{siunitx}
\usepackage[version=4]{mhchem}

\title{a package for modelling scattering-type scanning near-field optical microscopy}

\usepackage{orcidlink}

\usepackage{authblk}
\author[1,2]{Tom~Vincent\,\orcidlink{0000-0001-5974-9137}}
\author[1]{Xinyun~Liu\,\orcidlink{0000-0002-4275-7197}}
\author[1,2]{Daniel~Johnson\,\orcidlink{0000-0003-1655-3809}}
\author[3]{Lars~Mester\,\orcidlink{0000-0002-7094-3720}}
\author[2]{Nathaniel~Huang\,\orcidlink{0000-0001-7876-704X}}
\author[2,1]{Olga~Kazakova\,\orcidlink{0000-0002-8473-2414}}
\author[4,5]{Rainer~Hillenbrand\,\orcidlink{0000-0002-1904-4551}}
\author[1,2]{Jessica~Louise~Boland\,\orcidlink{0000-0002-6351-5699}}
\affil[1]{Photon Science Institute, Department of Electrical and Electronic Engineering, University~of~Manchester, Manchester, M13 9PL, UK}
\affil[2]{National Physical Laboratory, Hampton Road, Teddington, TW11 0LW, UK}
\affil[3]{Attocube Systems AG, Eglfinger Weg 2, 85540, Munich-Haar, Germany}
\affil[4]{CIC NanoGUNE BRTA and Department of Electricity and Electronics, Tolosa Hiribidea, Donostia-San Sebastián 20018, Spain}
\affil[5]{IKERBASQUE, Basque Foundation for Science, Bilbao 48011, Spain}

\usepackage[acronym]{glossaries}
\glsdisablehyper

\makeglossaries

\newacronym{afm}{AFM}{atomic force microscopy}
\newacronym{fdm}{FDM}{finite dipole model}
\newacronym{ftir}{FTIR}{Fourier transform infrared}
\newacronym{ir}{IR}{infrared}
\newacronym{lrm}{LRM}{lightning rod model}
\newacronym{pdm}{PDM}{point dipole model}
\newacronym{pmma}{PMMA}{poly(methyl methacrylate)}
\newacronym{ps}{PS}{poly(styrene)}
\newacronym{tmm}{TMM}{transfer matrix method}
\newacronym{to}{TO}{transverse optical}
\newacronym{snom}{s-SNOM}{scattering-type scanning near-field optical microscopy}

\newglossaryentry{i}
{
    name={\ensuremath{i}},
    description={imaginary unit},
}
\newglossaryentry{exp}
{
    name={\ensuremath{e}},
    description={Euler's number},
}
\newglossaryentry{j}
{
    name={\ensuremath{j}},
    description={index variable},
}
\newglossaryentry{m}
{
    name={\ensuremath{m}},
    description={index variable},
}
\newglossaryentry{X}
{
    name={\ensuremath{u}},
    description={integration substitution variable},
}
\newglossaryentry{f}
{
    name={\ensuremath{G}},
    description={arbitrary function},
}

\newglossaryentry{c}
{
    name={\ensuremath{c_0}},
    description={vacuum speed of light},
}
\newglossaryentry{hbar}
{
    name={\ensuremath{\hbar}},
    description={reduced Planck constant},
}
\newglossaryentry{e}
{
    name={\ensuremath{Q_e}},
    description={electron charge},
}

\newglossaryentry{x}
{
    name={\ensuremath{x}},
    description={spatial dimension parallel to sample plane},
}
\newglossaryentry{y}
{
    name={\ensuremath{y}},
    description={spatial dimension parallel to sample plane},
}
\newglossaryentry{z}
{
    name={\ensuremath{z}},
    description={spatial dimension perpendicular to sample plane},
}

\newglossaryentry{nu}
{
    name={\ensuremath{\nu}},
    description={vacuum wavenumber},
}
\newglossaryentry{q}
{
    name={\ensuremath{q}},
    description={in-plane wavevector component},
}
\newglossaryentry{omega}
{
    name={\ensuremath{\omega}},
    description={angular frequency},
}

\newglossaryentry{eps}
{
    name={\ensuremath{\varepsilon}},
    description={relative permitivitty},
}
\newglossaryentry{eps 0}
{
    name={\ensuremath{\gls{eps}_{0}}},
    description={vacuum permitivitty},
}
\newglossaryentry{eps inf}
{
    name={\ensuremath{\gls{eps}_{\infty}}},
    description={high-frequency permitivitty},
}
\newglossaryentry{eps sphere}
{
    name={\ensuremath{\gls{eps}_{tip}}},
    description={tip permitivitty},
}
\newglossaryentry{eps env}
{
    name={\ensuremath{\gls{eps}_{env}}},
    description={environment permitivitty},
}
\newglossaryentry{eps sub}
{
    name={\ensuremath{\gls{eps}_{sub}}},
    description={substrate permitivitty},
}

\newglossaryentry{lorentz}
{
    name={\ensuremath{\gls{eps}_{L}}},
    description={Lorentzian oscillator permittivity},
}
\newglossaryentry{drude}
{
    name={\ensuremath{\gls{eps}_{D}}},
    description={Drude function permittivity},
}
\newglossaryentry{gauss}
{
    name={\ensuremath{K_{G}}},
    description={Gaussian kernel},
}
\newglossaryentry{std}
{
    name={\ensuremath{\sigma_{G}}},
    description={Gaussian standard deviation},
}
\newglossaryentry{A osc}
{
    name={\ensuremath{A_{\gls{j}}}},
    description={Lorentzian oscillator strength},
}
\newglossaryentry{nu osc}
{
name={\ensuremath{\gls{nu}_{\gls{j}}}},
description={Lorentzian oscillator wavenumber},
}
\newglossaryentry{gamma}
{
    name={\ensuremath{\gamma}},
    description={carrier damping frequency},
}
\newglossaryentry{gamma osc}
{
name={\ensuremath{\gls{gamma}_{\gls{j}}}},
description={oscillator carrier damping frequency},
}
\newglossaryentry{nu plasma}
{
    name={\ensuremath{\gls{nu}_{p}}},
    description={plasma wavenumber},
}

\newcommand{\afm}[1]{\ensuremath{{#1}_{tip}}}
\newglossaryentry{z tip}
{
    name={\ensuremath{\afm{\gls{z}}}},
    description={\acrshort{afm} tip-sample separation},
}
\newglossaryentry{A tip}
{
    name={\ensuremath{\afm{A}}},
    description={\acrshort{afm} tapping amplitude},
}
\newglossaryentry{r tip}
{
    name={\ensuremath{\afm{r}}},
    description={\acrshort{afm} tip apex radius of curvature},
}
\newglossaryentry{omega tip}
{
    name={\ensuremath{\afm{\Omega}}},
    description={\acrshort{afm} tapping frequency},
}
\newglossaryentry{L tip}
{
    name={\ensuremath{\afm{L}}},
    description={\acrshort{fdm} ellipsoid semi-major axis length},
}

\newglossaryentry{potential}
{
    name={\ensuremath{\varphi}},
    description={electric potential},
}
\newglossaryentry{E s}
{
    name={\ensuremath{E_s}},
    description={scattered electric field strength},
}
\newglossaryentry{E z}
{
    name={\ensuremath{E_z}},
    description={$z$-component of electric field},
}
\newglossaryentry{E bg}
{
    name={\ensuremath{E_{bg}}},
    description={background electric field strength},
}
\newglossaryentry{E in}
{
    name={\ensuremath{E_{in}}},
    description={incident electric field strength},
}

\newglossaryentry{n}
{
    name={\ensuremath{n}},
    description={harmonic},
}
\newglossaryentry{demod op}
{
    name={\ensuremath{\mathop{\hat{\mathrm{F}}}}},
    description={tip demodulation operator},
}
\newglossaryentry{dirac op}
{
    name={\ensuremath{\mathop{\delta}}},
    description={dirac delta operator},
}

\newglossaryentry{sigma}
{
    name={\ensuremath{\sigma}},
    description={near-field scattering constant},
}
\newglossaryentry{sigma n}
{
name={\ensuremath{\gls{sigma}_{\gls{n}}}},
description={$\gls{n}^{\mathrm{th}}$-harmonic demodulated \glsdesc{sigma}},
}

\newglossaryentry{eta}
{
    name={\ensuremath{\eta}},
    description={\acrshort{snom} contrast},
}
\newglossaryentry{eta n}
{
name={\ensuremath{\gls{eta}_{\gls{n}}}},
description={$\gls{n}^{\mathrm{th}}$-harmonic demodulated \glsdesc{eta}},
}

\newglossaryentry{s}
{
    name={\ensuremath{s}},
    description={\acrshort{snom} amplitude},
}
\newglossaryentry{s n}
{
name={\ensuremath{\gls{s}_{\gls{n}}}},
description={$\gls{n}^{\mathrm{th}}$-harmonic demodulated \glsdesc{s}},
}

\newglossaryentry{phi}
{
    name={\ensuremath{\phi}},
    description={\acrshort{snom} phase},
}
\newglossaryentry{phi n}
{
name={\ensuremath{\gls{phi}_{\gls{n}}}},
description={$\gls{n}^{\mathrm{th}}$-harmonic demodulated \glsdesc{phi}},
}

\newglossaryentry{eff pol}
{
    name={\ensuremath{\alpha_{\mathrm{eff}}}},
    description={effective polarizability},
}
\newglossaryentry{eff pol n}
{
name={\ensuremath{\alpha_{\mathrm{eff}, \gls{n}}}},
description={$\gls{n}^{\mathrm{th}}$-harmonic demodulated \glsdesc{eff pol}},
}

\newglossaryentry{taylor coef}
{
name={\ensuremath{a_{\gls{j}, \gls{n}}}},
description={$\gls{j}^{\mathrm{th}}$ Taylor coefficient of \glsdesc{eff pol n}},
}
\newglossaryentry{max taylor}
{
    name={\ensuremath{J}},
    description={Maximum degree for a truncated Taylor expansion},
}

\newglossaryentry{r}
{
    name={\ensuremath{r}},
    description={Fresnel reflection coefficient},
}
\newglossaryentry{r p}
{
    name={\ensuremath{\gls{r}_{p}}},
    description={p-polarized \glsdesc{r}},
}
\newglossaryentry{r s}
{
    name={\ensuremath{\gls{r}_{s}}},
    description={s-polarized \glsdesc{r}},
}
\newglossaryentry{beta}
{
    name={\ensuremath{\beta}},
    description={quasistatic reflection coefficient},
}
\newglossaryentry{beta multi}
{
name={\ensuremath{\gls{beta}^{\gls{multi}}}},
description={multilayer effective quasistatic reflection coefficient},
}
\newglossaryentry{beta Q ave}
{
    name={\ensuremath{\overline{\gls{beta}}}},
    description={Charge-average-method definition of the effective quasistatic reflection coefficient},
}
\newglossaryentry{d}
{
    name={\ensuremath{t}},
    description={layer thickness},
}

\newglossaryentry{g}
{
    name={\ensuremath{g}},
    description={\acrshort{fdm} g factor},
}
\newglossaryentry{Q}
{
    name={\ensuremath{Q}},
    description={charge},
}
\newglossaryentry{Q0}
{
    name={\ensuremath{\gls{Q}_0}},
    description={charge 0},
}
\newglossaryentry{Q1}
{
    name={\ensuremath{\gls{Q}_1}},
    description={charge 1},
}
\newglossaryentry{Qj}
{
name={\ensuremath{\gls{Q}_{\gls{j}}}},
description={arbitrary charge},
}
\newglossaryentry{Qa}
{
    name={\ensuremath{\gls{Q}_{a}}},
    description={single representative charge},
}
\newglossaryentry{z Q}
{
name={\ensuremath{\gls{z}_{\gls{Q}}}},
description={height of a generic charge \gls{Q}},
}
\newglossaryentry{z Q0}
{
name={\ensuremath{\gls{z}_{\gls{Q0}}}},
description={height of a charge \gls{Q0}},
}
\newglossaryentry{z Q1}
{
name={\ensuremath{\gls{z}_{\gls{Q1}}}},
description={height of a charge \gls{Q1}},
}
\newglossaryentry{z Qj}
{
name={\ensuremath{\gls{z}_{\gls{Qj}}}},
description={height of a charge \gls{Qj}},
}
\newglossaryentry{z Qa}
{
name={\ensuremath{\gls{z}_{\gls{Qa}}}},
description={height of a charge \gls{Qa}},
}
\newglossaryentry{depth}
{
    name={\ensuremath{d}},
    description={depth below the surface},
}
\newglossaryentry{d Qj}
{
name={\ensuremath{\gls{depth}_{\gls{Qj}'}}},
description={depth of an image of charge \gls{Qj}},
}
\newglossaryentry{beta Q0}
{
name={\ensuremath{\gls{beta}_{\gls{Q0}'}}},
description={effective reflection coefficient for charge \gls{Qj}},
}
\newglossaryentry{beta Q1}
{
name={\ensuremath{\gls{beta}_{\gls{Q1}'}}},
description={effective reflection coefficient for charge \gls{Qj}},
}
\newglossaryentry{beta Qj}
{
name={\ensuremath{\gls{beta}_{\gls{Qj}'}}},
description={effective reflection coefficient for charge \gls{Qj}},
}

\newglossaryentry{p0}
{
    name={\ensuremath{p_0}},
    description={\gls{fdm} dipole 0},
}
\newglossaryentry{p1}
{
    name={\ensuremath{p_1}},
    description={\gls{fdm} dipole 1},
}

\newglossaryentry{geom}
{
    name={\ensuremath{f}},
    description={\gls{fdm} geometry function},
}

\newglossaryentry{bulk}
{
    name={\ensuremath{\mathrm{(B)}}},
    description={symbol denoting bulk sample},
}
\newglossaryentry{test}
{
    name={\ensuremath{\mathrm{(test)}}},
    description={symbol denoting test value},
}
\newglossaryentry{obs}
{
    name={\ensuremath{\mathrm{(obs)}}},
    description={symbol denoting observed value},
}
\newglossaryentry{geom bulk}
{
name={\ensuremath{\gls{geom}^{\gls{bulk}}}},
description={bulk \gls{fdm} geometry function},
}

\newglossaryentry{multi}
{
    name={\ensuremath{\mathrm{(M)}}},
    description={symbol denoting multilayer sample},
}
\newglossaryentry{geom multi}
{
name={\ensuremath{\gls{geom}^{\gls{multi}}}},
description={multilayer \gls{fdm} geometry function},
}

\newglossaryentry{alpha sphere}
{
    name={\ensuremath{\alpha_{tip}}},
    description={polarizability of a dielectric sphere representing the tip},
}
\newglossaryentry{point}
{
    name={\ensuremath{\mathrm{(P)}}},
    description={symbol denoting point dipole model},
}
\newglossaryentry{geom point}
{
name={\ensuremath{\gls{geom}^{\gls{point}}}},
description={point dipole model \gls{fdm} geometry function},
}

\newglossaryentry{lambda p}
{
    name={\ensuremath{\lambda_p}},
    description={polariton wavelength},
}
\newglossaryentry{tau}
{
    name={\ensuremath{\tau}},
    description={carrier relaxation time},
}
\newglossaryentry{E f}
{
    name={\ensuremath{E_f}},
    description={Fermi energy},
}
\newglossaryentry{cond}
{
    name={\ensuremath{\sigma_{gr}}},
    description={sheet conductivity of graphene},
}
\newglossaryentry{cond0}
{
    name={\ensuremath{\sigma_{0}}},
    description={universal high-frequency sheet conductivity of graphene},
}

\newglossaryentry{L1}
{
    name={\ensuremath{A}},
    description={initial layer},
}
\newglossaryentry{L2}
{
    name={\ensuremath{B}},
    description={subsequent layer},
}
\newglossaryentry{fwd}
{
    name={\ensuremath{R}},
    description={forward travelling wave},
}
\newglossaryentry{bwd}
{
    name={\ensuremath{L}},
    description={backwards travelling wave},
}
\newglossaryentry{T uv}
{
    name={\ensuremath{T_{\gls{L1}\gls{L2}}}},
    description={interface transfer matrix},
}
\newglossaryentry{P u}
{
    name={\ensuremath{P_{\gls{L1}}}},
    description={layer propagation matrix},
}
\newglossaryentry{ref}
{
    name={\ensuremath{\mathrm{(ref)}}},
    description={symbol denoting reference sample for normalization},
}

\newcommand{\demod}[1]{\gls{demod op}\left[#1\right]}
\newcommand{\dirac}[1]{\gls{dirac op}\left(#1\right)}
\newcommand{\caplet}[1]{\textbf{(#1)}}
\newcommand{\textlet}[1]{(#1)}

\newcommand{\docsurl}{https://snompy.readthedocs.io/}
\newcommand{\repourl}{https://github.com/TomVincentUK/snompy}

\makeatletter
\renewcommand{\maketitle}{%
  \begin{center}%
    \vskip 2\baselineskip\relax
    \includegraphics[width=0.8\linewidth]{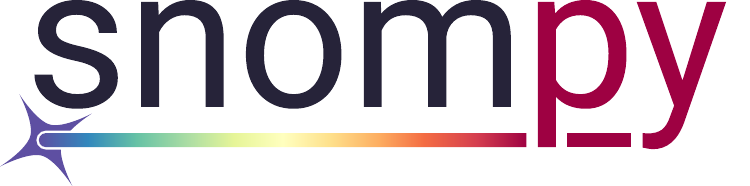}
    \par
    \vspace{-0.5\baselineskip}
    {\Huge\color{primary}\@title\par}
    \vskip 2\baselineskip\relax
    {\large\@author\par}
\end{center}%
}
\makeatother

\begin{document}

\twocolumn[
    \begin{@twocolumnfalse}
        \maketitle

        \vskip \baselineskip\relax

        \begin{tabularx}{\linewidth}{>{\centering\arraybackslash}X >{\centering\arraybackslash}X}
            \hline
            \href{\docsurl}{\Large{Documentation}} & \href{\repourl}{\Large{Code repository}} \\
            \hline
        \end{tabularx}

        \vskip 2\baselineskip\relax

        \begin{abstract}
            \Gls{snom} is a powerful technique for extreme subwavelength imaging and spectroscopy, with \SI{\sim 20}{\nano m} spatial resolution.
            But quantitative relationships between experiment and material properties requires modelling, which can be computationally and conceptually challenging.
            In this work, we present snompy an open-source Python library which contains implementations of two of the most common \gls{snom} models, the \gls{fdm} and the \gls{pdm}.
            We show a series of typical uses for this package with demonstrations including simulating nano-\gls{ftir} spectra and recovering permittivity from experimental \gls{snom} data.
            We also discuss the challenges faced with this sort of modelling, such as competing descriptions of the models in literature, and finite size effects.
            We hope that snompy will make quantitative \gls{snom} modelling more accessible to the wider research community, which will further empower the use of \gls{snom} for investigating nanoscale material properties.
        \end{abstract}

    \end{@twocolumnfalse}
]

\clearpage
\section{Introduction}

\begin{figure*}
    \centering
    \includegraphics{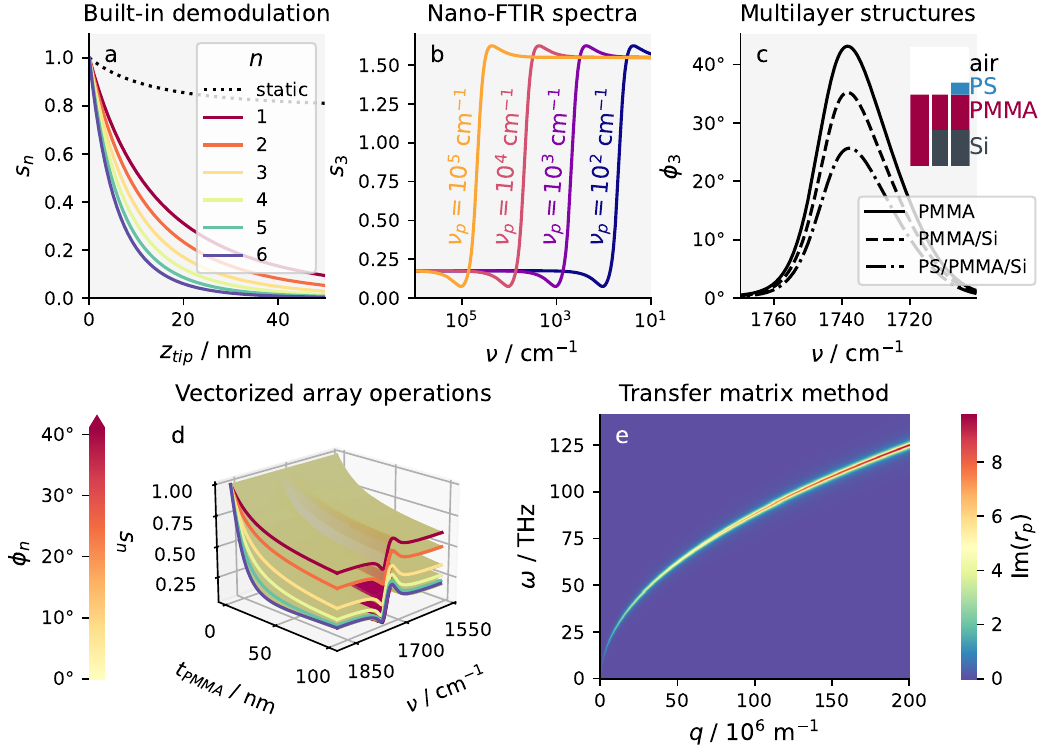}
    \caption{Showcase of snompy capabilities.
    \caplet{a} Modulation and demodulation at harmonics, \gls{n}, of an \acrfull{afm} tip's oscillation. This example uses the \acrfull{pdm} to simulate the normalized decay of the \gls{snom} amplitude, \gls{s n}, with tip-sample separation, \gls{z tip}, for the static case and for several demodulation orders.
    \caplet{b} Nano-\gls{ftir} spectra.
    This example uses the \acrfull{fdm} to simulate third-harmonic amplitude spectra as a function of wavenumber, \gls{nu}, from hypothetical semiconducting materials with varying plasma frequency, \gls{nu plasma}.
    \caplet{c} Multilayer samples.
    This example uses the multilayer \acrfull{fdm} to simulate third-harmonic phase spectra from bulk \gls{pmma}, \SI{60}{\nano m} \gls{pmma}/\ce{Si}, and \SI{20}{\nano m} \gls{ps}/\SI{60}{\nano m} \gls{pmma}/\ce{Si}.
    \caplet{d} Vectorized operations using NumPy-style broadcasting \autocite{Harris2020}.
    This example uses the multilayer \acrfull{fdm} to simulate \gls{snom} spectra from \gls{pmma}/\ce{Si} for a range of \gls{pmma} thicknesses, and demodulation harmonics.
    All \num{183906} points shown were calculated simultaneously with no Python loops.
    The edges of the 3d surfaces for each harmonic are coloured to match \textlet{a}.
    \caplet{e} \Acrfull{tmm} for simulating far-field (Fresnel) reflection coefficients.
    This example shows plasmonic dispersion for a sample of graphene on \ce{Si} via the imaginary part of the in-plane wavevector-, \gls{q}-, dependent Fresnel reflection coefficient, \gls{r p}.
    The permittivity values used for \ce{Si} and \gls{ps}  \autocite{Mester2020} are \(\gls{eps}_{\ce{Si}}=11.7\) and \(\gls{eps}_{\textrm{\gls{ps}}}=2.5\).
    The permittivity model used for the hypothetical semiconductors is detailed in section \ref{sec:eps_semi}.
    The permittivity model used for \gls{pmma} is shown in figure \ref{fig:spectra}\textlet{c} and detailed in section \ref{sec:eps_pmma}.
    The permittivity model used for graphene is detailed in section \ref{sec:eps_graphene}.
    All \gls{s n} and \gls{phi n} values are normalized to a bulk \ce{Si} reference.
    The specific parameters used for \textlet{a-d} are also summarized in table \ref{tab:settings}.}
    \label{fig:greatest_hits}
\end{figure*}

\Acrfull{snom} is a well-established technique for probing optical properties of materials at length scales far smaller than the diffraction limit.
It spans a wide spectrum across the
visible \autocite{Hu2017,Chen2011},
near-\gls{ir} \autocite{Hu2017b},
mid-\gls{ir} \autocite{Fei2012,Vincent2020,Greaves2023,Kaltenecker2021,Dai2015,Chen2020,Sunku2018,Mester2020,Mester2022},
and \si{\tera\hertz} \autocite{Mooshammer2018,Plankl2021,Chen2021}
regions, and has applications ranging from
chemical nanoidentification \autocite{Mester2020,Greaves2023},
to 2D \autocite{Vincent2020,Hu2017,Hu2017b,Fei2012,Plankl2021,Dai2015,Sunku2018}
and topological \autocite{Mooshammer2018,Chen2021}
materials,
to plasmonics and metamaterials \autocite{Dai2015,Chen2011,Fei2012,Plankl2021,Chen2020,Sunku2018},
to the biological sciences \autocite{Greaves2023,Kaltenecker2021}.

But, despite its prevalence, interpretation of \gls{snom} data remains challenging and requires modelling of the electromagnetic interaction between the sample and an \gls{afm} tip under illumination.
Various models, such as the \gls{pdm} \autocite{Keilmann2004}, \gls{fdm} \autocite{Cvitkovic2007,Hauer2012,Mester2020,Mester2020b}, and \gls{lrm} \autocite{McLeod2014}, have been proposed, which show varying levels of quantitative agreement with experimental results.
However these models can be difficult to understand, and tricky to implement with computational efficiency.

To address this issue, we have created \emph{snompy}, an open-source Python package that provides implementations of the \gls{fdm} and \gls{pdm}, as well as a set of related tools for modelling \gls{snom} data.
Figure \ref{fig:greatest_hits} demonstrates some of the main use cases and functionalities of snompy, which we list below.

A crucial feature is the ability to simulate lock-in amplifier measurements of \gls{snom} signals via modulation and demodulation at arbitrary harmonics, \gls{n}, of an \gls{afm} tip's oscillation, as shown by the approach curves in figure \ref{fig:greatest_hits}\textlet{a}.

A typical use of snompy, as demonstrated in figure \ref{fig:greatest_hits}\textlet{b} is to simulate nano-\gls{ftir} spectra from user-input dielectric permittivities, \gls{eps}, which vary according to wavenumber, \gls{nu}.
In this example, the changing third-harmonic \gls{snom} amplitude spectra from a set of hypothetical semiconducting materials with varying plasma frequency, \gls{nu plasma}, are simulated.
The permittivities are modelled using a Drude function as detailed in section \ref{sec:eps_semi}.

We have also introduced a new way to evaluate quasistatic reflection coefficients from layered structures using an adapted form of the \gls{tmm}, describe in section \ref{sec:tmm}, which enables simulations of \gls{snom} measurements from samples with an arbitrary number of layers.
This is shown by the simulated nano-\gls{ftir} spectra from several multilayer samples featuring \gls{pmma}, \gls{ps} and \ce{Si}, shown in figure \ref{fig:greatest_hits}\textlet{c}.

\Gls{snom} modelling can be computationally intensive, particularly for multilayer samples, as it relies on evaluating both definite and semi-definite integrals \autocite{Hauer2012,Mester2020, Mester2020b}.
We address this challenge in snompy by using Gauss-Laguerre quadrature to convert semi-definite integrals to finite sums, as described in section \ref{sec:integrals}.
This allows us to improve efficiency by taking advantage of NumPy-style, vectorized array broadcasting \autocite{Harris2020}.
It also eliminates the need to import third-party quadrature routines, meaning snompy's only dependencies are Python and NumPy.

Figure \ref{fig:greatest_hits}\textlet{d} is an example simulation that takes advantage of these optimizations.
It shows a series of simulated nano-\gls{ftir} spectra from \gls{pmma} on \ce{Si} with three independent variables: \gls{pmma} thickness, \(\gls{d}_{\mathrm{\gls{pmma}}}\); wavenumber, \gls{nu}; and demodulation order, \gls{n}.
The \num{183906} resulting points shown were calculated using vectorized array operations with no need for high-level Python loops.

Although \gls{snom} is a near-field technique, far-field (Fresnel) reflection coefficients are an important part of \gls{snom} modelling, as discussed in section \ref{sec:background}.
As such, snompy features an implementation of the far-field \gls{tmm}, which can calculate Fresnel reflection coefficients from structures with an arbitrary number of layers.
This functionality can be used in isolation from the \gls{fdm} and \gls{pdm}, for example to simulate polariton dispersion in polariton interferometry experiments \autocite{Chen2021,Dai2015,Fei2012,Hu2017,Hu2017b}.
To demonstrate this, figure \ref{fig:greatest_hits}\textlet{e} shows a simulation of the plasmonic dispersion of a layer of graphene on \ce{Si}. The permittivity model used for graphene is detailed in section \ref{sec:eps_graphene}.

In the rest of this paper, we will describe the theoretical background needed to understand \gls{snom} modelling in section \ref{sec:background}.
We will then the \gls{fdm} and \gls{pdm} as they are implemented in snompy in sections \ref{sec:fdm} and \ref{sec:pdm}.
Section \ref{sec:results} shows several demonstrations and discussions of key uses for snompy: simulating thin-film nano-\gls{ftir} spectra in section \ref{sec:spectra}, the quantitative effects of varying model and model parameters in section \ref{sec:params}, and recovering permittivity from \gls{snom} data in section \ref{sec:taylor}.

The final part of this paper, section \ref{sec:methods} describes some of the important implementation details of snompy, as well as details of the permittivity models are used in this paper.
For a comprehensive look at the details of snompy, along with tutorials and example scripts, we point users to the \href{\docsurl}{online documentation}.

\subsection{Background}
\label{sec:background}

At the heart of the modelling approach taken by snompy is the fundamental equation governing the near-field interaction in scattering \gls{snom} \autocite{Cvitkovic2007}:
\begin{equation}
    \gls{sigma} \equiv \frac{\gls{E s}}{\gls{E in}} = (1 + \gls{r p})^2 \gls{eff pol}
    \label{eq:fundamental}
\end{equation}
Here \gls{sigma} is the scattering coefficient, which is the signal of interest, and is defined as the ratio of the incident and tip-scattered light's electric fields, \gls{E in} and \gls{E s}. The quantity \gls{r p} is the far-field Fresnel reflection coefficient.

\begin{figure}
    \centering
    \includegraphics{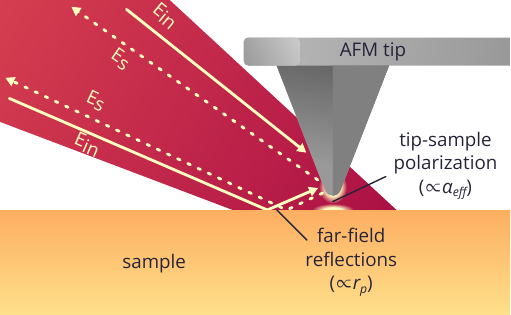}
    \caption{Diagram showing how an illuminated \gls{afm} tip scatters near-field-confined light back into the far field.}
    \label{fig:tip_sample}
\end{figure}

The other terms of this equation can be understood by referring to figure \ref{fig:tip_sample}, which shows a typical scattering \gls{snom} experiment.
A focused light beam with incident electric field \gls{E in} shines onto an \gls{afm} tip and sample, which induces an instantaneous near-field charge polarization between the tip's apex and sample's surface, whose magnitude is governed by the tip-sample effective polarizability, \gls{eff pol}.
This polarization scatters light back into the far field with electric field \gls{E s}, meaning \(\gls{sigma} \propto \gls{eff pol}\), which shows that \gls{eff pol} contains the information about the near-field part of the \gls{snom} signal.

The efficiency of scattering is enhanced by the elongated probe of the typically metal-coated \gls{afm} tip acting as an antenna, via the lightning-rod effect \autocite{Keilmann2004}.

Calculating \gls{eff pol} requires modelling of the nanoscale light-matter interactions between the \gls{afm} tip and sample.
It is strongly governed by the sample's electric permittivity \gls{eps} (and thickness in the case of layered samples), or equivalently the quasistatic reflection coefficient of the sample's surface \gls{beta}.
In snompy, effective polarizability can be calculated via the \gls{fdm} or \gls{pdm}, as described in sections \ref{sec:fdm} and \ref{sec:pdm}.

The \gls{afm} tip is illuminated both directly, and indirectly by reflection from the sample surface with reflection coefficient \gls{r p} (for p-polarized light).
Similarly, the scattered light from the tip is also detected directly, and indirectly after reflection from the surface.
This is accounted for by the far-field reflection term \((1 + \gls{r p})^2\), which doesn't contain near-field information, but still modifies the detected \gls{snom} signal.\footnotemark{}
Snompy features an implementation of the \gls{tmm} \autocite{Zhan2013} for calculating far-field (Fresnel) reflection coefficients, which is described in section \ref{sec:tmm}.

\footnotetext{This factor is occasionally written as \((1 + c \gls{r p})^2\), where \(c\) is an empirical weighting factor that represents the finite proportion of reflected light that takes part in the interaction \autocite{Aizpurua2008,Mester2022}.}

In experiments, the amplitude of the background scattering from the far-field light spot is many orders of magnitude greater than the amplitude of the near-field-scattered light.
So scattering \gls{snom} experiments typically operate the \gls{afm} in tapping mode, then use a lock-in amplifier to demodulate the detected signal at high harmonics, \gls{n}, of the tapping frequency.
The detected signal is then given, not by equation \ref{eq:fundamental} but by  \autocite{Keilmann2004}:
\begin{equation}
    \gls{sigma n} = (1 + \gls{r p})^2 \gls{eff pol n}
    \label{eq:demod}
\end{equation}

Here, and throughout this paper, a subscript \gls{n} refers to quantities which are demodulated at the \(\gls{n}^{\mathrm{th}}\) harmonic of the \gls{afm} tapping frequency.
For details of how these quantities are calculated, see section \ref{sec:demod}.

A final nuance that needs to be accounted for is that the measured \gls{sigma n} signal depends on instrumental factors, such as the detector sensitivity and \gls{E in}, which are usually unknown.
For quantitative results, it is therefore necessary to normalize the measured values to measurements taken from a reference sample with a well-known dielectric function.
Au and Si are common choices as they are spectrally flat in the mid-\gls{ir}.

The normalized signal, often called the near-field contrast \autocite{Cvitkovic2007,Mester2022}, is therefore given by:
\begin{equation}
    \gls{eta n}
    = \frac{\gls{sigma n}}{\gls{sigma n}^{\gls{ref}}}
    = \frac{(1 + \gls{r p})^2}{\left(1 + \gls{r p}^{\gls{ref}}\right)^2}
    \left(\frac{\gls{eff pol n}}{\gls{eff pol n}^{\gls{ref}}}\right)
    \label{eq:contrast}
\end{equation}

Here a superscript \gls{ref} refers to parameters taken from the reference sample used for normalization.

The parameter \gls{eta n} is a complex number that represents both the amplitude, \gls{s n}, and phase, \gls{phi n}, of the near-field scattered light as:
\begin{equation}
    \begin{aligned}
        \gls{eta n} = & \gls{s n} \gls{exp}^{\gls{i} \gls{phi n}}  \\
        \gls{s n} =   & \left|\gls{eta n}\right|                   \\
        \gls{phi n} = & \operatorname{arg}\left(\gls{eta n}\right)
    \end{aligned}
    \label{eq:s_phi}
\end{equation}

\subsection{Finite dipole model}
\label{sec:fdm}

\begin{figure}
    \centering
    \includegraphics{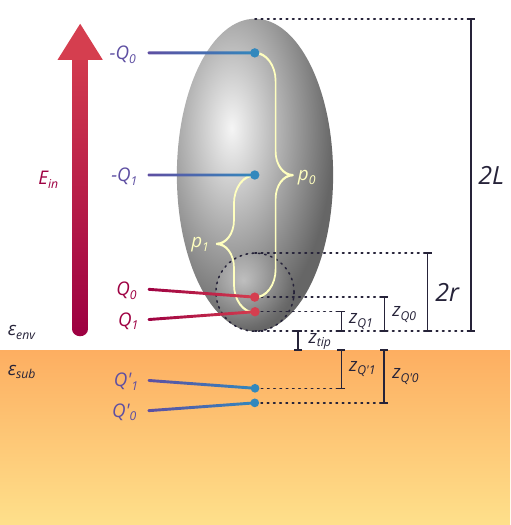}
    \caption{Image showing an ellipsoid model of an \gls{afm} tip over a semi-infinite sample, along with the model charges and image charges used to derive the bulk \gls{fdm}.}
    \label{fig:fdm}
\end{figure}

The \gls{fdm} was initially derived for bulk samples by Cvitkovic \textit{et al.} (2007) \autocite{Cvitkovic2007}, then extended to multilayer samples by Hauer \textit{et al.} (2012) \autocite{Hauer2012}.
A slightly modified implementation was also presented by Mester \textit{et al.} (2020) \autocite{Mester2020}.

These are the versions included in snompy, and thus described below, however we stress that there are several other \gls{fdm} implementations described in literature \autocite{Govyadinov2013,Govyadinov2014,Barnett2020}.

\subsubsection{Bulk samples}
\label{sec:bulk}

The \gls{fdm} models an \gls{afm} tip as a perfectly conducting ellipsoid, with radius of curvature \gls{r tip} and semi-major axis length \gls{L tip}, as shown in figure \ref{fig:fdm}.
By making the quasistatic approximation, we can model the response of the tip to far-field light by the response to a vertically oriented static electric field of strength \gls{E in}.

In the absence of a sample, the field around the ellipsoid can be approximated by a dipole formed of two charges \(\pm \gls{Q0}\) at distances \gls{z Q0} from the ends of the ellipsoid \autocite{Ocelic2007}, where:
\begin{equation}
    \gls{z Q0} \approx \frac{1.31 \gls{L tip} \gls{r tip}}{\gls{L tip} + 2 \gls{r tip}}
    \label{eq:z_Q0}
\end{equation}

The strength of this dipole is given by:
\begin{equation}
    \begin{aligned}
        \gls{p0} = & 2 (\gls{L tip} - \gls{z Q0}) \gls{Q0}                                 \\
        \approx    & 2 \gls{L tip} \gls{Q0}, \quad \text{for } \gls{r tip} \ll \gls{L tip}
    \end{aligned}
    \label{eq:p0}
\end{equation}

When the tip is brought close to a sample surface, we make the assumption that only the charge \gls{Q0} interacts with the sample, due to the large distance from the \(-\gls{Q0}\) charge.
The tip-sample interaction can then be modelled using the method of image charges.

The response of the sample to \gls{Q0} can be modelled with an image charge \(\gls{Q0}'\) below the interface.
For bulk samples, the value of this charge is determined by the quasistatic reflection coefficient, \gls{beta}, as:
\begin{equation}
    Q_{0}' = \gls{beta} \gls{Q0}
    \label{eq:Q0_image}
\end{equation}

The value of \gls{beta} can be found from the environment and substrate permittivities \gls{eps env} and \gls{eps sub}, as:
\begin{equation}
    \gls{beta} = \frac{\gls{eps env} - \gls{eps sub}}{\gls{eps env} + \gls{eps sub}}
    \label{eq:beta}
\end{equation}

This image also asserts an influence back on the tip, which we model with an extra charge \gls{Q1} at a distance \(\gls{z Q1}=\frac{\gls{r tip}}{2}\) from the end of the tip.
\gls{Q1} has its own image, \(\gls{Q1}'\), and for charge conservation we must add a counter charge \(-\gls{Q1}\) at the centre of the tip \autocite{Cvitkovic2007,Hauer2012}.

The value of \gls{Q1} can be solved for by accounting for contributions to the overall polarization from \gls{Q0}, and also from \gls{Q1} itself, as:
\begin{equation}
    \gls{Q1} = \gls{beta}(\gls{geom bulk}_0 Q_0 + \gls{geom bulk}_1 \gls{Q1})
    \label{eq:Q1}
\end{equation}

Here the \(\gls{geom bulk}_{\gls{j}}\) values are functions encapsulating the geometry of the system\footnotemark{}, given by:
\begin{equation}
    \begin{aligned}
         & \gls{geom bulk}_{\gls{j}} =                                                                                                                                                                                                         \\
         & \left(\gls{g} - \frac{\gls{r tip} + 2 \gls{z tip} + \gls{z Qj}}{2 \gls{L tip}}\right)\frac{\ln{\left(\frac{4 \gls{L tip}}{\gls{r tip} + 4 \gls{z tip} + 2 \gls{z Qj}}\right)}}{\ln{\left(\frac{4 \gls{L tip}}{\gls{r tip}}\right)}}
    \end{aligned}
    \label{eq:geom_func_bulk}
\end{equation}

The factor \gls{g} here is a dimensionless weighting factor that modifies the proportion of the total charge that takes part in the interaction, this is typically set to \(\gls{g}=0.7 \gls{exp}^{0.06 \gls{i}}\).

\footnotetext{We use the superscript \gls{bulk} to distinguish these bulk geometry functions \gls{geom bulk} from their multilayer equivalents \gls{geom multi} used in the multilayer implementation of the \gls{fdm} (see section \ref{sec:multilayer}).}

The charges \gls{Q1} and \(-\gls{Q1}\) form another dipole:
\begin{equation}
    \begin{aligned}
        \gls{p1} = & (\gls{L tip} - \gls{z Q1}) \gls{Q1}                                 \\
        \approx    & \gls{L tip} \gls{Q1}, \quad \text{for } \gls{r tip} \ll \gls{L tip}
    \end{aligned}
    \label{eq:p1}
\end{equation}

The effective polarizability can then be found from the total induced dipole, as:
\begin{equation}
    \begin{aligned}
        \gls{eff pol} = & \frac{\gls{p0} + \gls{p1}}{\gls{E in}}                                                                                               \\
        \approx         & 2 \gls{L tip} \frac{\gls{Q0}}{\gls{E in}} \left(1 + \frac{\gls{geom bulk}_0 \gls{beta}}{2 (1 - \gls{geom bulk}_1 \gls{beta})}\right)
    \end{aligned}
    \label{eq:with_prop}
\end{equation}

The constant of proportionality here can be neglected as it depends on the unquantified \(\frac{\gls{Q0}}{\gls{E in}}\) ratio, leading to the final result:
\begin{equation}
    \gls{eff pol} \propto 1 + \frac{\gls{geom bulk}_0 \gls{beta}}{2 \left(1 - \gls{geom bulk}_1 \gls{beta}\right)}
    \label{eq:bulk}
\end{equation}

\subsubsection{Multilayer samples}
\label{sec:multilayer}

Hauer \textit{et al.} proposed a method to extend the \gls{fdm} to multilayer samples \autocite{Hauer2012}.
In this method, the response of the multilayer sample to a charge \gls{Qj} is modelled with a single image charge \(\gls{Qj}'=\gls{beta Qj}\gls{Qj}\) at a depth \gls{d Qj} below the sample surface.

The effective depth and reflection coefficient for the charge are chosen to be those that preserve the normal electric field \gls{E z} and potential \gls{phi} at the surface of the sample, as:
\begin{equation}
    \begin{aligned}
        \gls{d Qj}    & = \left|\frac{\left.\gls{potential}\right|_{\gls{z}=0}}{\left.\gls{E z}\right|_{\gls{z}=0}}\right| - \gls{z Qj}, \quad \mathrm{and} \\
        \gls{beta Qj} & = \frac{\left(\left.\gls{potential}\right|_{\gls{z}=0}\right)^2}{\left.\gls{E z}\right|_{\gls{z}=0}}
    \end{aligned}
    \label{eq:hauer bcs}
\end{equation}

The potential and field can be calculated from:

\begin{equation}
    \begin{aligned}
        \left.\gls{potential}\right|_{\gls{z}=0} & = \int_0^{\infty} \gls{beta multi}(\gls{q}) e^{-2 \gls{z Qj} \gls{q}} d\gls{q}, \quad \mathrm{and} \\
        \left.\gls{E z}\right|_{\gls{z}=0}       & = \int_0^{\infty} \gls{beta multi}(\gls{q}) \gls{q} e^{-2 \gls{z Qj} \gls{q}} d\gls{q}
    \end{aligned}
    \label{eq:phi_E}
\end{equation}

Here \gls{q} is the in-plane momentum of light, and \gls{beta multi} is the effective quasistatic reflection coefficient for the surface (see section \ref{sec:tmm} for details of how this is calculated in snompy).

These values can then be inserted into a modified version of equation \ref{eq:bulk}, as:

\begin{equation}
    \gls{eff pol} \propto 1 + \frac{\gls{geom multi}_0 \gls{beta Q0}}{2 \left(1 - \gls{geom multi}_1 \gls{beta Q1}\right)}
    \label{eq:hauer}
\end{equation}

The geometry function is also modified in this case to become:

\begin{equation}
    \begin{aligned}
         & \gls{geom multi}_{\gls{j}} =                                                                                                                                                                                                      \\
         & \left(\gls{g} - \frac{\gls{r tip} + \gls{z tip} + \gls{d Qj}}{2 \gls{L tip}}\right)\frac{\ln{\left(\frac{4 \gls{L tip}}{\gls{r tip} + 2 \gls{z tip} + 2 \gls{d Qj}}\right)}}{\ln{\left(\frac{4 \gls{L tip}}{\gls{r tip}}\right)}}
    \end{aligned}
    \label{eq:geom_func_multi}
\end{equation}

\subsubsection{Charge average method}
\label{sec:Q_ave}

Mester \textit{et al.} proposed another multilayer \gls{fdm} method \autocite{Mester2020}, which is also included in snompy.

In this implementation, the geometry function used is \gls{geom bulk}, the same as for the bulk \gls{fdm} method, however an alternative expression for the quasistatic reflection coefficient is used, which is derived from the ratio of the fields at the height of the probe:

\begin{equation}
    \gls{beta Q ave} = \frac{\int_0^{\infty} \gls{beta multi}(\gls{q}) \gls{q} e^{-2 \gls{z Qa} \gls{q}} d\gls{q}}
    {\int_0^{\infty} \gls{q} e^{-2 \gls{z Qa} \gls{q}} d\gls{q}}
    \label{eq:beta_mester}
\end{equation}

The height \gls{z Qa} here is the height of a single representative test charge \gls{Qa}, whose position within the tip is chosen empirically.
A value of \(\gls{z Qa} = \gls{z tip} + 1.4 \gls{r tip}\) was found to produce the best agreement with experimental \gls{snom} spectra in the original reference \autocite{Mester2020}.

This can then be inserted into another modified version of equation \ref{eq:bulk}, as:

\begin{equation}
    \gls{eff pol} \propto 1 + \frac{\gls{geom bulk}_0 \gls{beta Q ave}}{2 \left(1 - \gls{geom bulk}_1 \gls{beta Q ave}\right)}
    \label{eq:Q ave}
\end{equation}

\subsection{Point dipole model}
\label{sec:pdm}

The \gls{pdm} is an older alternative to the \gls{fdm} which typically has less quantitative agreement with experiment \autocite{Cvitkovic2007,Hauer2012}.
However, it has some advantages in that it is conceptually simpler and can account for the dielectric function of the \gls{afm} tip, so it is also featured in snompy.

In the \gls{pdm}, the tip is first modelled as a dielectric sphere whose response to electric fields is then represented by a single point dipole, located in the sphere's centre \autocite{Keilmann2004,Cvitkovic2007}.
The effective polarizability is given by:

\begin{equation}
    \gls{eff pol} = \frac{\gls{alpha sphere}}{1 - \gls{geom point} \gls{beta}}
    \label{eq:pdm}
\end{equation}

Here \gls{alpha sphere} represents the polarizability of the model sphere, which relates to the tip permittivity \gls{eps sphere} and radius \gls{r tip} as:

\begin{equation}
    \gls{alpha sphere} = 4 \pi \gls{r tip}^3 \frac{\gls{eps sphere} - 1}{\gls{eps sphere} + 2}
    \label{eq:alpha sphere}
\end{equation}

The function \gls{geom point} encapsulates extra geometric properties of the system and is given by:

\begin{equation}
    \gls{geom point} = \frac{\gls{alpha sphere}}{16 \pi \left(\gls{r tip} + \gls{z tip}\right)^3}
    \label{eq:geom point}
\end{equation}

\section{Results}
\label{sec:results}

\subsection{Thin-film nano-\glsentrytext{ftir} spectra}
\label{sec:spectra}

\begin{figure}
    \centering
    \includegraphics{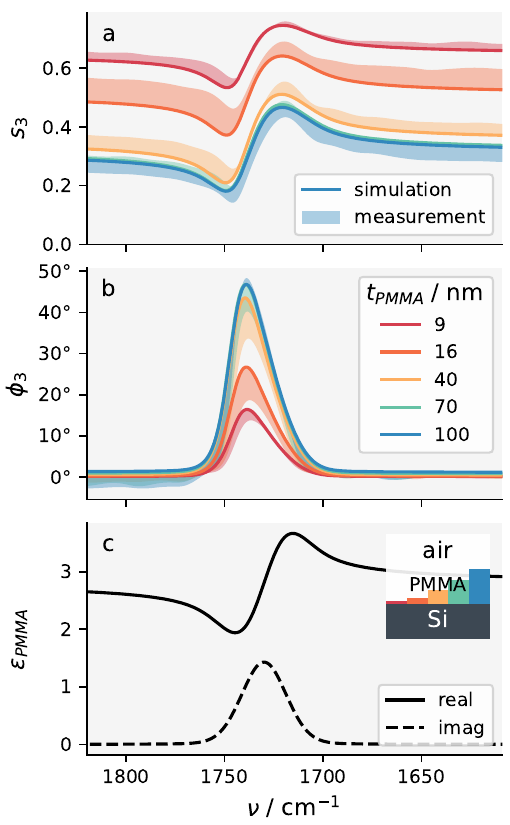}
    \caption{
    Simulated and experimental nano-\gls{ftir} spectra from samples with varying thicknesses of \gls{pmma} on \ce{Si}.
    \caplet{a} The third-harmonic \gls{snom} amplitude.
    \caplet{b} The third-harmonic \gls{snom} phase.
    \caplet{c} The simulated dielectric function of \gls{pmma}, chosen to best reproduce the experimental data (inset: schematic of sample layer structure).
    Filled curves in \textlet{a} and \textlet{b} show the differences between the experimental and simulated data.
    The experimental data is reproduced with permission from Mester \textit{et al.} \autocite{Mester2020}.
    All signals shown are normalized to a bulk \ce{Si} reference with \(\gls{eps}_{\ce{Si}}=11.7\).
    The specific parameters used for each plot are also summarized in table \ref{tab:settings}.
    }
    \label{fig:spectra}
\end{figure}

In this section we will demonstrate a typical use of snompy by using the \gls{fdm} to simulate nano-\gls{ftir} spectra from different thicknesses of \gls{pmma} on a \ce{Si} substrate, normalized against a bulk \ce{Si} reference, and to compare them with experimental data reproduced with permission from Mester \textit{et al.} \autocite{Mester2020}.

Figures \ref{fig:spectra}\textlet{a} and \textlet{b} show simulated and experimental third-harmonic amplitude, \(\gls{s}_3\), and phase, \(\gls{phi}_3\), spectra from \gls{pmma} samples ranging from \num{9} to \SI{100}{\nano m} in thickness.

To simulate a \gls{snom} signal using the \gls{fdm}, we need to input various parameters which describe the effective shape and properties of the \gls{afm} tip.
For this simulation we used the parameters reported in the original work by Mester \textit{et al.}, and these parameters are reported in table \ref{tab:settings}.
The effects of various parameters of the \gls{fdm} are discussed in section \ref{sec:params}.

We also need to input dielectric permittivities for all of the constituent materials.
In the mid-\gls{ir}, the permittivity of \ce{Si} is given by \(\gls{eps}_{\ce{Si}}=11.7\)  \autocite{Mester2020}.
We represent the permittivity of \gls{pmma}, \(\gls{eps}_{\textrm{\gls{pmma}}}\), by a broadened Lorentzian oscillator at the frequency of its \ce{C=O} bond.
This is shown in figure \ref{fig:spectra}\textlet{c}.
The specific parameters for \(\gls{eps}_{\textrm{\gls{pmma}}}\) were chosen to best reproduce the experimental data, and are given in section \ref{sec:eps_pmma}.

Note that the resulting permittivity shown is slightly higher than previous measurements acquired by ellipsometry from bulk samples \autocite{Mester2020}.
This could potentially arise from sample differences, such as different proportions of cross-linking between bulk and thin-film samples.

However the permittivity we show is qualitatively similar to previous measurements, and leads to a good agreement between simulation and experiment, so we conclude that it is a reasonable model for the \gls{pmma} in these measurements.

\subsection{Effects of varying model and parameters}
\label{sec:params}

\begin{figure*}
    \centering
    \includegraphics{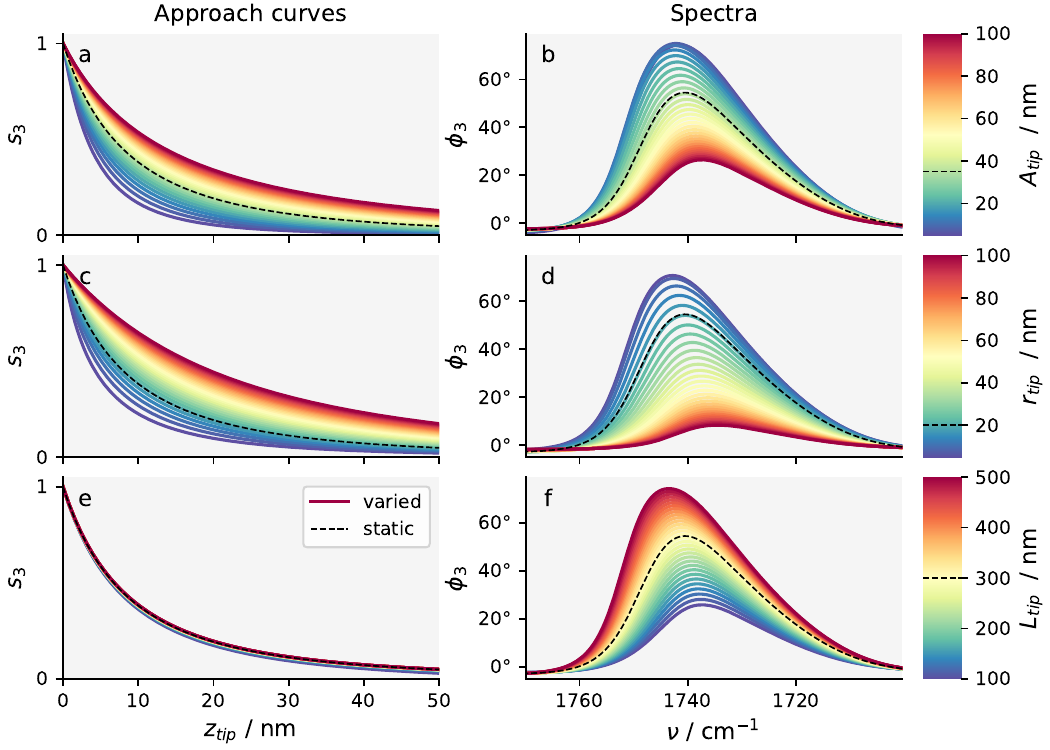}
    \caption{
    Effects of varying model parameters.
    \Gls{fdm}-simulated approach curves from a bulk \ce{Si} sample and spectra from a sample of \SI{60}{\nano m} of \gls{pmma} on \ce{Si} with varying: \caplet{a, b} tapping amplitude, \gls{A tip}; \caplet{c, d} tip radius, \gls{r tip}; and \caplet{e, f} tip semi-major axis length, \gls{L tip}.
    Black dashed lines on colour scales represent the fixed value of the corresponding quantity used in the other plots of this figure.
    The dielectric permittivity model used for \gls{pmma} is shown in figure \ref{fig:spectra}\textlet{c}.
    All signals shown are normalized to a bulk \ce{Si} reference with \(\gls{eps}_{\ce{Si}}=11.7\).
    The specific parameters used for each plot are also summarized in table \ref{tab:settings}.
    }
    \label{fig:approach_curves}
\end{figure*}

For quantitative comparisons between modelling and experiments, it is important to choose model parameters reproduce the measured results well.
A common way to calibrate the model parameters is to fit them to experimental approach curves \autocite{Hauer2012,Cvitkovic2007,Lupo2021}, in which the signal is measured as a function of tip-sample separation, \gls{z tip}.

As described in section \ref{sec:fdm}, the \gls{fdm} represents an \gls{afm} tip as a perfectly conducting ellipsoid with tapping amplitude \gls{A tip}, apex radius of curvature \gls{r tip}, and semi-major axis length \gls{L tip}.
In this section we use the \gls{fdm} to investigate the effects of varying these parameters on a series of approach curves taken from \ce{Si}, with permittivity \(\gls{eps}_{\ce{Si}} = 11.7\), and a series of spectra taken from a sample of \SI{60}{\nano m} of \gls{pmma} on \ce{Si}, where the modelled permittivity of \gls{pmma} is the same as shown in figure \ref{fig:spectra}\textlet{c}.

Figures \ref{fig:approach_curves}\textlet{a} and \textlet{b} show the effect of varying the tapping amplitude.
The approach curves reveal that a lower tapping amplitude leads to a faster decay of the \gls{snom} signal with distance from the sample, \gls{z tip}.
This is because a reduced tapping amplitude leads to a reduced interaction volume, which consequently increases the surface sensitivity of the measurement \autocite{Mooshammer2018}, meaning tapping amplitude can be an effective way to adjust the degree of near-field confinement.

Figure \ref{fig:approach_curves}\textlet{b} demonstrates that tapping amplitude also has an effect on the lineshape and apparent frequency of resonances in nano-\gls{ftir} spectra.
This shows that correctly calibrated measurements of tapping amplitude are important for quantitative \gls{snom} modelling.

Figures \ref{fig:approach_curves}\textlet{c} to \textlet{f} show the effects of varying the effective radius and length of the model ellipsoid.
Unlike tapping amplitude, which is typically a controlled parameter in \gls{snom} experiments, these may be unknown values which require fitting to best reproduce the experimental data.

The spectra reveal that both tip radius and length have an effect on the amplitude and apparent frequency of resonances in nano-\gls{ftir} spectra, however in opposite directions.
Comparatively, the approach curves demonstrate that the radius has a more significant effect on the decay of the near-field signal.

\begin{figure*}
    \centering
    \includegraphics{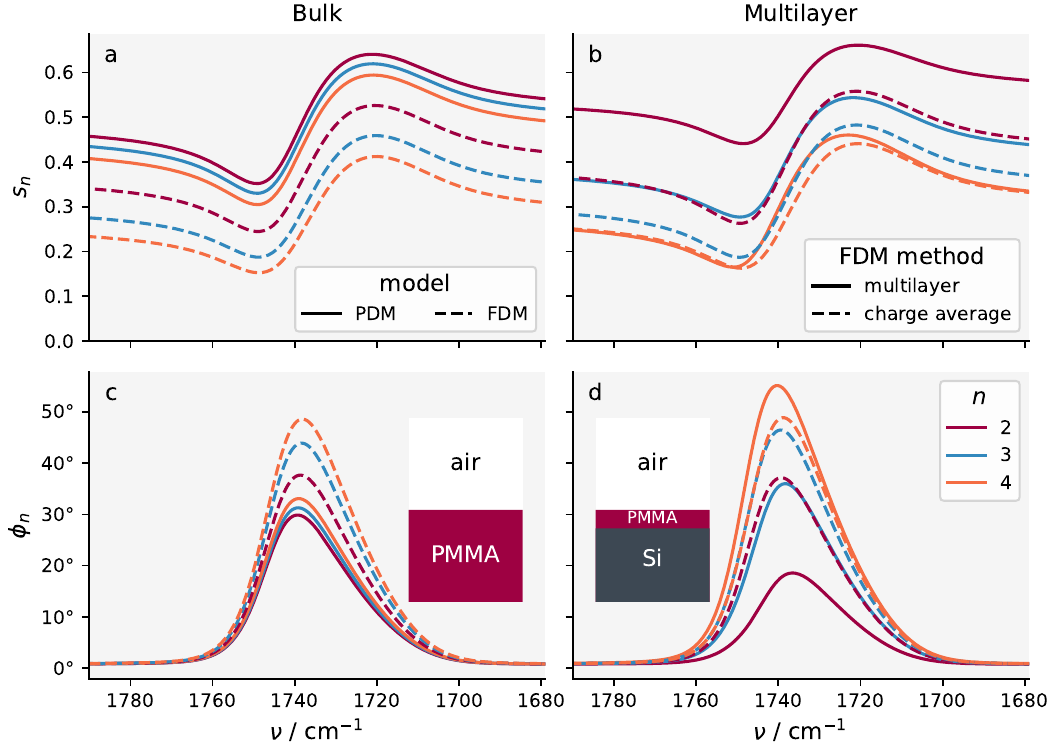}
    \caption{
    Quantitative differences between different \gls{snom} models.
    The spectra on the left compare the \acrfull{pdm} and \acrfull{fdm} for a sample of bulk \gls{pmma}, and the spectra on the right compare two different multilayer \gls{fdm} implementations for a sample of \SI{60}{\nano m} \gls{pmma} on bulk \ce{Si}.
    Panels show: \caplet{a, b} amplitude spectra, \gls{s n}; and \caplet{c, d} phase spectra, \gls{phi n}, for three demodulation harmonics, \gls{n}.
    Insets in \textlet{c} and \textlet{d} show schematics of the simulated materials.
    The dielectric permittivity model used for \gls{pmma} is shown in figure \ref{fig:spectra}\textlet{c}.
    All signals shown are normalized to a bulk \ce{Si} reference with \(\gls{eps}_{\ce{Si}}=11.7\).
    The specific parameters used for each plot are also summarized in table \ref{tab:settings}.
    }
    \label{fig:differences}
\end{figure*}

As well as the effects of varying parameters within a particular model, it is also important to consider the effects of changing the model itself.
Figure \ref{fig:differences} highlights the quantitative differences between simulations using the \gls{pdm} and \gls{fdm} with similar parameters, and also between the multilayer model as described by Hauer \textit{et al.} (see section \ref{sec:multilayer}) and the charge average method (see section \ref{sec:Q_ave}).

The results in this section highlight the importance of well-chosen model and model parameters for quantitative agreement with experiment.
As such, it is important that any modelling tool allows full and flexible control over the constituent parameters of the models it implements.
Section \ref{sec:api} details the main functions included in snompy and gives descriptions of all controllable parameters.

\subsection{Recovering material properties}

The typical use case for the \gls{fdm}, as demonstrated above, is to simulate \gls{snom} measurements from user-input model samples.
Comparitively few works have focused on using the \gls{fdm} to recover sample properties from experimental \gls{snom} measurements \autocite{Govyadinov2013,Govyadinov2014,Mooshammer2018,Lupo2021}, though this is also an attractive use case.

In this section we will discuss some different approaches to tackle this problem, as well as some typical challenges encountered in the process.

Sample properties in the \gls{fdm} can be represented via their permittivity \gls{eps}.
The primary challenge in inverting the \gls{fdm}, is that the expression for the \gls{snom} contrast, \gls{eta n}, as a function of \gls{eps} is impossible to invert algebraically \autocite{Govyadinov2013,Govyadinov2014} (see sections \ref{sec:background} and \ref{sec:fdm} for a description of this function).

\begin{figure*}
    \centering
    \includegraphics{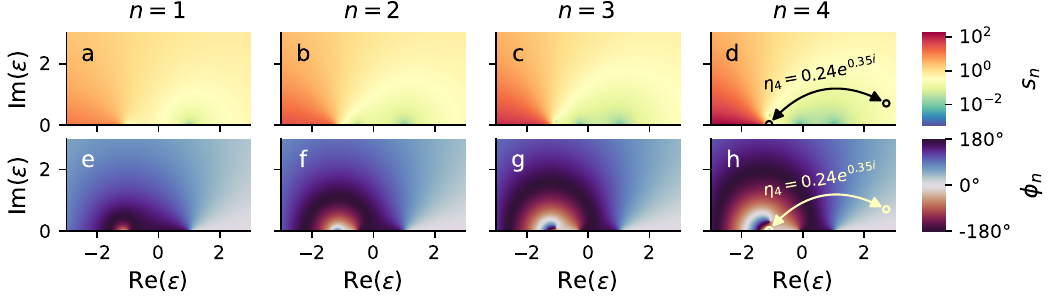}
    \caption{
        \Gls{snom} signals as a function of permittivity.
        Images show: \caplet{a-d} the amplitude, \gls{s n}; and \caplet{e-f} the phase ,\gls{phi n}; of a \gls{snom} measurement normalized to bulk \ce{Si} for harmonics \(\gls{n}=1\) to \num{4}.
        Note the log scaling of the colour mapping in figures \textlet{a-d}.
        Arrows and small circles in panels \textlet{d} and \textlet{h} show an example of two points with different \gls{eps} values but indistinguishable \(\gls{eta}_4 = \gls{s}_4 e^{i \gls{phi}_4}\) values.
        The specific parameters used for this calculation are summarized in table \ref{tab:settings}.
    }
    \label{fig:field}
\end{figure*}

To better visualize the problem we are trying to solve, figure \ref{fig:field} plots \(\gls{eta n} = \gls{s n} e^{i \gls{phi n}}\), as a function of \gls{eps} for four harmonics, \gls{n}, normalized to a \ce{Si} reference with \(\gls{eps}_{\ce{Si}}=11.7\).
This reveals the complicated variation of \gls{eta n}, with some regions appearing relatively flat while others feature lots of local variation (note the log scaling of the \gls{s n} scale).
The varying sensitivity of \gls{eta n} to small changes in \gls{eps} can make recovering \gls{eps} challenging, particularly in the presence of experimental noise, as small errors in \gls{eta n} can lead to relatively large errors in the corresponding \gls{eps} value.

One approach to recovering \gls{eps} that has had some success in literature \autocite{Mooshammer2018,Lupo2021}, is to use minimization routines to minimize the difference between the observed and simulated \gls{eta n}.
However, this approach can be computationally expensive, as it typically relies on many independent function calls.

It also suffers from the fact that \gls{eta n} as a function of \gls{eps} is a poorly conditioned function for minimization.
For example, on top of the above-mentioned varying sensitivity of the function, there are also multiple values of \gls{eps} which can produce identical values of \gls{eta n}.
One such pair of values is indicated as an example in figures \ref{fig:field}\textlet{d} and \textlet{h}.
These indistinguishable points can potentially lead to false solutions during minimization.

The number of false solutions can be reduced by restricting the permitted \gls{eps} values to only those with a positive imaginary part, as we have done here.

\begin{figure*}
    \centering
    \includegraphics{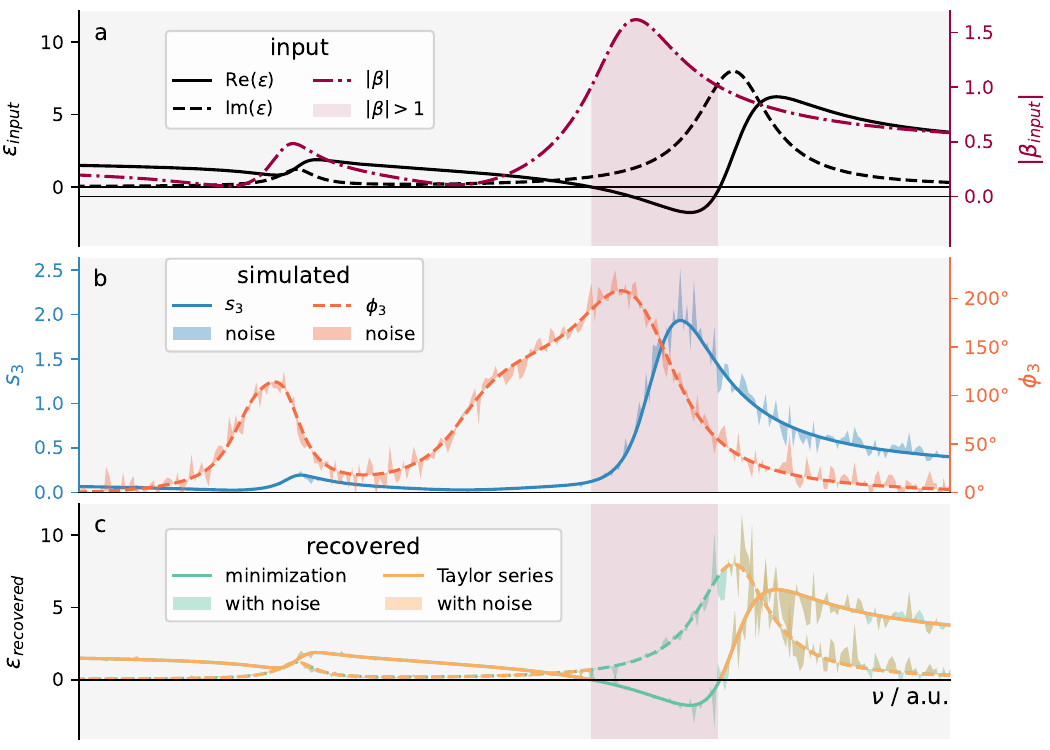}
    \caption{
        Recovering permittivity from \gls{snom} signals.
        \caplet{a} A simulated permittivity, \gls{eps}, and its corresponding quasistatic reflection coefficient, \gls{beta}, featuring both a weak oscillator (\(|\gls{beta}| < 1\)) and a strong oscillator (\(|\gls{beta}| > 1\)).
        The region of strong light-matter interaction  with \(|\gls{beta}| > 1\) is coloured in red.
        \caplet{b} Simulated third-harmonic \gls{snom} signal \(\gls{eta}_3 = \gls{s}_3 e^{i \gls{phi}_3}\), using the \gls{fdm}.
        Some simulated experimental noise was added, shown by filled areas above and below the lines.
        \caplet{c} Permittivity recovered from the pristine and noisy \gls{snom} spectra shown in \textlet{b}, using both minimization and by using the Taylor series method.
        Note that the Taylor series method returns no valid solutions in the region where \(|\gls{beta}| > 1\).
        The specific parameters used for these calculations are summarized in table \ref{tab:settings}.
    }
    \label{fig:taylor_inversion}
\end{figure*}

We show an example of a simulated experiment that uses minimization to attempt to recover \gls{eps} from \gls{eta} in figure \ref{fig:taylor_inversion}.
A simulated permittivity, shown in figure \ref{fig:taylor_inversion}\textlet{a}, is used to generate a third-harmonic nano-\gls{ftir} spectrum, shown in figure \ref{fig:taylor_inversion}\textlet{b}.

SciPy's \verb|scipy.optimize.minimize| function \autocite{Virtanen2020} is then used to find the \gls{eps} value that minimizes the function:

\begin{equation}
    \left|\gls{eta}_3^{(\gls{test})}(\gls{eps}) - \gls{eta}_3^{(\gls{obs})}\right|
    \label{eq:minimization}
\end{equation}

Here a superscript \gls{obs} refers to the observed value and \gls{test} to the \gls{eps}-dependent test value that should be adjusted to match it.
The results from this are shown by the dashed lines in figure \ref{fig:taylor_inversion}\textlet{c}.
To show the sensitivity of the recovery to noise, some simulated noise is added to the spectra in figure \ref{fig:taylor_inversion}\textlet{b}, shown by filled areas above and below the curves, and the results of minimizing the noisy spectra are also shown in figure \ref{fig:taylor_inversion}\textlet{c}.

Another successful approach \autocite{Govyadinov2013,Govyadinov2014,Mester2020} is to express \gls{eff pol n} as a truncated Taylor series in \gls{beta}, which can be related to \gls{eps} using equation \ref{eq:beta}.
The expression is then possible to invert algebraically and solutions can be found by finding the complex roots of a high-order polynomial.
The derivation of the Taylor expansion for the \gls{fdm} used in snompy is shown in section \ref{sec:taylor}.

A downside of this approach is that the Taylor expansion only converges for values of \(|\gls{beta}| < \sim 1\).
This is usually only true for weak molecular oscillators, which prevents the method being used for samples with strong light-matter oscillations, such as metals and polaritonic materials.

This Taylor series method of inversion is implemented natively in snompy, with no need to import external minimization routines as we did for the example above.
We ran this inversion on the same simulated spectra as for the previous example and the results from this are shown by the solid lines in figure \ref{fig:taylor_inversion}\textlet{c}.
Note that there are no solutions returned in the coloured area of the graph, where \(|\gls{beta}| > 1\).

We stress that the recovery of \gls{eps} from the simulated \gls{snom} spectra in this section are an idealized case, which is free from systematic errors and in which the same model is used to both generate the data and to perform the recovery.
When working with experimental data, the limitations and applicability of the model must be taken into account.
Indeed, one final and fundamental challenge to inverting the \gls{fdm} to recover material properties from experimental data, is that the model does not account for finite-size effects \autocite{Cvitkovic2007, Mester2022}.
Samples are modelled as semi-infinite planar structures, which means any effects related to factors such as reflections or standing waves cannot be reproduced.
A detailed discussion of these effects is beyond the scope of this paper, but the extent to which finite size effects can be accounted for in \gls{fdm} modelling could form the basis of future studies.

\section{Discussion}

The models we have made accessible as open-source code in snompy have had considerable success in literature for producing quantitative agreement with experiments \autocite{Cvitkovic2007,Hauer2012,Mester2020}.
However, there are certain challenges which must also be addressed.

One particularly common issue is that these models assume semi-infinite, planar samples, so they cannot account for effects introduced by finite lateral sample dimensions, which includes many commonly studied effects such as surface polaritons and standing waves in nanophotonic structures.
This is particularly problematic, as one of the main motivators for using \gls{snom} instead of far-field optical microscopy is to image structures smaller than the diffraction limit, so finite lateral dimensions are a characteristic feature of many \gls{snom} samples.
Other models, which are not currently implemented in snompy, such as those described by Chen \textit{et al.}  \autocite{Chen2021b,Chen2022}, have attempted to address this by accounting for laterally varying sample dimensions.

Furthermore, in the process of adapting models from their descriptions in literature, we have identified many conflicting interpretations of the \gls{fdm} \autocite{Cvitkovic2007,Hauer2012,Govyadinov2013,Govyadinov2014,Mester2022,Barnett2020}, which differ in terms of how they account for multilayer samples, and how they evaluate quasistatic reflection coefficients, among other details.
This is a natural consequence of different groups working in parallel on complicated models.

We showed in section \ref{sec:params} that two models included in snompy, the Hauer multilayer method \autocite{Hauer2012} and the charge average method \autocite{Mester2020,Mester2020b}, lead to quantitative differences for comparable input parameters, and we can expect that other implementations will produce similar model-to-model quantitative differences.
We stress that this is not an indication that any particular model is at fault, only that the parameters needed to agree with experimental data must be chosen for the specific implementation being used.
However, these discrepancies do warrant further comparison, and this is one area where we believe open-source modelling software can be particularly beneficial.

We note that of all the \gls{fdm} implementations we encountered, the description by Hauer \textit{et al.} \autocite{Hauer2012} seems to be the most ubiquitous in literature, so we have chosen this as the default implementation in snompy.

The extent to which these challenges interfere with quantitative \gls{snom} modelling must be understood before this modelling can become a widely accessible and useful tool for probing nanoscale material properties.
But this will require methodical work, and comparison between models and carefully designed test samples to identify the regimes where the models are no longer applicable.
By publishing snompy, which contains efficient and usable open-source implementations of the \gls{fdm}, we hope to make this vital work faster, easier and more transparent.

A further advantage of making snompy open-source is that we can invite contributions from the wider research community, and take advantage of expertise beyond that of our own group and collaborators.
For example, in future versions of snompy, we would be keen to include additional \gls{snom} models, beyond the \gls{fdm} and \gls{pdm} which we have already implemented, such as the \gls{lrm} \autocite{McLeod2014} and those models that account for varying sample geometry \autocite{Chen2021b,Chen2022}.
In this way, we hope that snompy will be a valuable and evolving resource for the \gls{snom} community for many years to come.

\section{Methods}
\label{sec:methods}

\subsection{Programming interface details}
\label{sec:api}

In this section we give some details of the programming interface to the functions contained within snompy, we also give the specific parameters used for the example simulations in this paper.
We stress that this section is intended as an overview, and is not a guide for users of snompy.
For a comprehensive look at the details of snompy, along with tutorials and example scripts, we point users to the \href{\docsurl}{online documentation}.

The primary methods provided by snompy are the expressions for effective polarizability, \gls{eff pol}, and the \(\gls{n}^{th}\)-harmonic-demoduated effective polarizability, \gls{eff pol n}, using the \gls{fdm} and \gls{pdm}, which are summarized below in table \ref{tab:primary_methods}.

\begin{table}[H]
    \centering
    \begin{tabularx}{\columnwidth}{c c l}
        model     & parameter       & method                        \\
        \hline
        \gls{fdm} & \gls{eff pol}   & \verb|snompy.fdm.eff_pol()|   \\
        \gls{fdm} & \gls{eff pol n} & \verb|snompy.fdm.eff_pol_n()| \\
        \gls{pdm} & \gls{eff pol}   & \verb|snompy.pdm.eff_pol()|   \\
        \gls{pdm} & \gls{eff pol n} & \verb|snompy.pdm.eff_pol_n()| \\
    \end{tabularx}
    \caption{Effective polarizability methods in snompy.}
    \label{tab:primary_methods}
\end{table}

Of these, the \verb|eff_pol_n()| functions are adaptations of the \verb|eff_pol()| functions with extra arguments related to demodulation.
We will list the main arguments for the two \verb|eff_pol_n()| functions to show the degree of control possible with snompy.
The arguments for \verb|snompy.fdm.eff_pol_n()| are given in table \ref{tab:fdm_args}, and the arguments for \verb|snompy.pdm.eff_pol_n()| are given in table \ref{tab:pdm_args}.

\begin{table*}
    \centering
    \begin{tabularx}{\textwidth}{c c X}
        argument        & symbol                             & description                                                                                                            \\
        \hline
        \verb|sample|   & |                                  & Object representing a sample.                                                                                          \\
        \verb|A_tip|    & \gls{A tip}                        & The tapping amplitude of the \gls{afm} tip.                                                                            \\
        \verb|n|        & \gls{n}                            & The harmonic of the \gls{afm} tip tapping frequency at which to demodulate.                                            \\
        \verb|z_tip|    & \gls{z tip}                        & Height of the tip above the sample.                                                                                    \\
        \verb|r_tip|    & \gls{r tip}                        & Radius of curvature of the \gls{afm} tip.                                                                              \\
        \verb|L_tip|    & \gls{L tip}                        & Semi-major axis length of the effective spheroid from the finite dipole model.                                         \\
        \verb|g_factor| & \gls{g}                            & Empirical weighting factor (described in section \ref{sec:bulk}).                                                      \\
        \verb|d_Q0|     & \(\frac{\gls{z Q0}}{\gls{r tip}}\) & Depth of an induced charge 0 within the tip, in units of \gls{r tip}.                                                  \\
        \verb|d_Q1|     & \(\frac{\gls{z Q1}}{\gls{r tip}}\) & Depth of an induced charge 1 within the tip, in units of \gls{r tip}.                                                  \\
        \verb|method|   & |                                  & The method of the finite dipole model to use: \verb|"bulk"|, \verb|"multi"| or \verb|"Q_ave"|.                         \\
        \verb|d_Qa|     & \(\frac{\gls{z Qa}}{\gls{r tip}}\) & Depth of a test charge \(a\) within the tip, in units of \gls{r tip} (only used by the \verb|"Q_ave"| implementation). \\
    \end{tabularx}
    \cprotect\caption{Arguments for \verb|snompy.fdm.eff_pol_n()|. The first column is the variable name in the code and the second is the mathematical symbol used in this paper.}
    \label{tab:fdm_args}
\end{table*}

\begin{table*}
    \centering
    \begin{tabularx}{\textwidth}{c c X}
        argument         & symbol             & description                                                                    \\
        \hline
        \verb|sample|    & |                  & Object representing a sample.                                                  \\
        \verb|A_tip|     & \gls{A tip}        & The tapping amplitude of the \gls{afm} tip.                                    \\
        \verb|n|         & \gls{n}            & The harmonic of the \gls{afm} tip tapping frequency at which to demodulate.    \\
        \verb|z_tip|     & \gls{z tip}        & Height of the tip above the sample.                                            \\
        \verb|r_tip|     & \gls{r tip}        & Radius of curvature of the \gls{afm} tip.                                      \\
        \verb|eps_tip|   & \gls{eps sphere}   & Permittivity of the tip (ignored if \verb|alpha_tip| specified).               \\
        \verb|alpha_tip| & \gls{alpha sphere} & Polarizability of the conducting sphere used as a model for the \gls{afm} tip. \\
    \end{tabularx}
    \cprotect\caption{Arguments for \verb|snompy.pdm.eff_pol_n()|. The first column is the variable name in the code and the second is the mathematical symbol used in this paper.}
    \label{tab:pdm_args}
\end{table*}

The first argument for both \verb|eff_pol_n()| should be an instance of a \verb|snompy.Sample| object.
This is an object that represents a sample as a stack of dielectric layers with a semi-infinite substrate and superstrate (the environment above the sample surface).
The user is able to specify the permittivity and thickness of each layer to define the \gls{snom} simulation.
The sample object also has methods for calculating far-field and quasistatic reflection coefficients as described in section \ref{sec:tmm}.

The specific parameters used to calculate all \gls{eta n} simulations shown in this work are summarized in table \ref{tab:settings}

\begin{table*}
    \centering
    \begin{tabularx}{\textwidth}{ c c c c c c c X }
        figure                                      & \gls{n}         & \gls{A tip} / \si{\nano m} & \gls{r tip} / \si{\nano m} & \gls{L tip} / \si{\nano m} & \gls{g}                          & \gls{eps sphere} & method                                  \\
        \hline
        \ref{fig:greatest_hits}\textlet{a}          & \numrange{1}{6} & \num{30}                   & \num{30}                   & |                          & |                                & \(\infty\)       & \gls{pdm}                               \\
        \ref{fig:greatest_hits}\textlet{b}          & \num{3}         & \num{30}                   & \num{30}                   & \num{200}                  & \num{0.6}                        & |                & \gls{fdm} (bulk)                        \\
        \ref{fig:greatest_hits}\textlet{c}          & \num{3}         & \num{30}                   & \num{30}                   & \num{200}                  & \num{0.6}                        & |                & \gls{fdm} (multilayer)                  \\
        \ref{fig:greatest_hits}\textlet{d}          & \numrange{1}{6} & \num{30}                   & \num{30}                   & \num{200}                  & \num{0.6}                        & |                & \gls{fdm} (multilayer)                  \\
        \ref{fig:greatest_hits}\textlet{d}          & \numrange{1}{6} & \num{30}                   & \num{30}                   & \num{200}                  & \num{0.6}                        & |                & \gls{fdm} (multilayer)                  \\
        \ref{fig:spectra}\textlet{a, b}             & \num{3}         & \num{30}                   & \num{30}                   & \num{200}                  & \num{0.6}                        & |                & \gls{fdm} (charge average)              \\
        \ref{fig:approach_curves}\textlet{a, c, e}* & \num{3}         & \num{35}                   & \num{20}                   & \num{300}                  & \(0.7 \gls{exp}^{0.06 \gls{i}}\) & |                & \gls{fdm} (bulk)                        \\
        \ref{fig:approach_curves}\textlet{b, d, f}* & \num{3}         & \num{35}                   & \num{20}                   & \num{300}                  & \(0.7 \gls{exp}^{0.06 \gls{i}}\) & |                & \gls{fdm} (multilayer)                  \\
        \ref{fig:differences}\textlet{a, c}         & \numrange{2}{4} & \num{30}                   & \num{30}                   & \num{200}                  & \num{0.6}                        & \(\infty\)       & \gls{pdm} / \gls{fdm} (bulk)            \\
        \ref{fig:approach_curves}\textlet{b, c}     & \numrange{2}{4} & \num{30}                   & \num{30}                   & \num{200}                  & \num{0.6}                        & |                & \gls{fdm} (multilayer / charge average) \\
        \ref{fig:field}                             & \numrange{1}{4} & \num{35}                   & \num{20}                   & \num{300}                  & \(0.7 \gls{exp}^{0.06 \gls{i}}\) & |                & \gls{fdm} (bulk)                        \\
        \ref{fig:taylor_inversion}\textlet{b}       & \num{3}         & \num{30}                   & \num{20}                   & \num{300}                  & \(0.7 \gls{exp}^{0.06 \gls{i}}\) & |                & \gls{fdm} (bulk)                        \\
    \end{tabularx}
    \caption{
    Input parameters for all \gls{eta n} calculations shown in this work.
    All \gls{eta n}, \gls{s n} and \gls{phi n} signals shown are normalized to a sample of bulk \ce{Si} with \(\gls{eps}_{\ce{Si}}=11.7\) acquired with the same model parameters.
    *Parameters for figure \ref{fig:approach_curves} represent the fixed values only, shown by black dashed lines in the plots.
    }
    \label{tab:settings}
\end{table*}

\subsection{Simulating demodulation}
\label{sec:demod}

\Gls{snom} relies on lock-in amplifier demodulation of \gls{z tip}-dependent signals at different harmonics, \gls{n}, of the \gls{afm} tip's tapping frequency.

To model this we first define a height modulation as:

\begin{equation}
    \gls{z tip}'(\theta; \gls{z tip}, \gls{A tip}) = \gls{z tip} + \gls{A tip} \left(1 + \cos{\left(\theta\right)}\right)
    \label{eq:z_mod_time}
\end{equation}

Here \(\gls{z tip}'\) is the modulated version of \gls{z tip}, \gls{A tip} is the tapping amplitude and \(\theta\) is the phase of the \gls{afm}'s tapping cycle\footnotemark{}.

\footnotetext{
    Note that the cosine term here is offset so that the bottom of the oscillation is at \gls{z tip} (rather than at \(\gls{z tip} - \gls{A tip}\)).
    This ensures that when the nominal value of \(\gls{z tip} = 0\) the oscillation never becomes negative (which would mean intersecting with the sample surface).
}

To simulate demodulation at a harmonic, \gls{n}, we multiply the input signal by a complex sinusoid, then integrate over one period of the tip oscillation.

We represent this with an operator \gls{demod op} to convert an arbitrary, \gls{z tip}-dependent function \gls{f} into its modulated and demodulated equivalent \(\gls{f}_n\), as:

\begin{equation}
    \begin{aligned}
         & \gls{f}_{\gls{n}}(\gls{z tip}; \gls{A tip}, \gls{n})                                                                                                   \\
         & \equiv \demod{\gls{f}(\gls{z tip}); \gls{A tip}, \gls{n}}                                                                                              \\
         & \equiv \frac{1}{2 \pi} \int_{-\pi}^{\pi} \gls{f}\left(\gls{z tip}'(\theta; \gls{z tip}, \gls{A tip})\right) \gls{exp}^{\gls{i} \gls{n} \theta} d\theta
    \end{aligned}
    \label{eq:demod_operator}
\end{equation}

This can be evaluated efficiently using simple quadrature methods such as the trapezium method.

\subsection{Taylor series representation}
\label{sec:taylor}

For samples with weak light-matter interactions, such that the magnitude of the quasistatic reflection coefficient \(|\gls{beta}| < \sim1\), \gls{beta} can be recovered from the demodulated effective polarizability \gls{eff pol n} with the help of a Taylor series representation of \gls{eff pol n} \autocite{Govyadinov2013,Govyadinov2014,Mester2020,Mester2020b}.
In this section we describe how this is implemented in snompy.

For the bulk \gls{fdm} formula given in equation \ref{eq:bulk} we can represent the effective polarizability by a Taylor series in \gls{beta} as:

\begin{equation}
    \gls{eff pol} \propto  1 + \frac{\gls{geom bulk}_0}{2} \sum_{\gls{j}=1}^{\infty} \left(\gls{geom bulk}_1\right)^{\gls{j} - 1} \gls{beta}^{\gls{j}}
    \label{eq:fdm_taylor}
\end{equation}

Using the notation defined in section \ref{sec:demod}, equation \ref{eq:fdm_taylor} can then be extended to higher harmonics of effective polarizability as:

\begin{equation}
    \gls{eff pol n}
    \propto \demod{1 + \frac{\gls{geom bulk}_0}{2} \sum_{\gls{j}=1}^{\infty} \left(\gls{geom bulk}_1\right)^{\gls{j} - 1} \gls{beta}^{\gls{j}}}
    \label{eq:fdm_taylor_demod}
\end{equation}

This can be expanded then rearranged to the form:

\begin{equation}
    \begin{aligned}
        \gls{eff pol n}   &
        \propto \dirac{\gls{n}} + \sum_{\gls{j}=1}^{\infty} \gls{taylor coef} \gls{beta}^{\gls{j}} \\
        \gls{taylor coef} &
        = \frac{\demod{\gls{geom bulk}_0 \left(\gls{geom bulk}_1\right)^{\gls{j} - 1}}}{2}
    \end{aligned}
    \label{eq:fdm_taylor_final}
\end{equation}

Here \(\dirac{n}\) represents the Dirac delta function, and \gls{taylor coef} represents the \(\gls{j}^\mathrm{th}\) Taylor coefficient.
This series converges for \(|\gls{beta}| < \sim1\), which is generally true for weak molecular oscillators.

Truncating the series in equation \ref{eq:fdm_taylor_final} to a finite number of terms, \gls{max taylor}, means that an unknown \gls{beta} can be recovered from a known \gls{eff pol n} by finding the roots of the resulting polynomial.

This typically leads to multiple solutions, as for a \(\gls{max taylor}^\mathrm{th}\)-order complex polynomial there may be up to \gls{max taylor} unique roots.
However, in practice, we find that these can usually be reduced to a single valid solution by rejecting solutions for which \(|\gls{beta}| > \sim1\) and optionally rejecting those for which \(\mathrm{Im}\left(\gls{eps}\right)<0\).

\subsection{Reflection coefficients}
\label{sec:tmm}

Accurate \gls{snom} modelling requires calculation of both far-field (electrodynamic) and near-field (quasistatic) reflection coefficients.
Multilayer \gls{fdm} methods also require calculation of the effective quasistatic reflection coefficient for layered structures \gls{beta multi}.

Previous works have derived explicit equations for \gls{beta multi} for particular numbers of layers, by repeated application of the method of image charges \autocite{Hauer2012,Lupo2021}, but this is computationally challenging and requires a quickly expanding number of terms for each extra layer.
Mester \textit{et al.} proposed a way to generalize this to an arbitrary number of layers by recursively applying the equation for a 3-layer structure \autocite{Mester2020}, but this fails for reflection coefficients close to zero (which can occur off-resonance in multilayer structures).

In snompy we instead adapt the \gls{tmm} to calculate effective quasistatic reflection coefficients.
This provides a computationally efficient method which works for arbitrary numbers of layers and has scope to be extended thanks to its relationship to the well-researched far-field \gls{tmm}.

In this section we describe how electrodynamic and quasistatic reflection coefficients are calculated in snompy.
We begin with a recap of the well-known electrodynamic \gls{tmm}, then move on to describe how we adapted it to work for the quasistatic regime.

\subsubsection{Electrodynamic}

The \gls{tmm} relates the field of electromagnetic waves on one side of a stack of interfaces to the other side.

In snompy we define that the interfaces are stacked, and that the fields propagate, in the \gls{z} direction.
Then for a layer \gls{L1}, we define field coefficients for forward and backwards travelling waves as \(\gls{fwd}_{\gls{L1}}\) and \(\gls{bwd}_{\gls{L1}}\), and we define the \gls{z}-component of the wavevector in the layer as \(\gls{nu}_{\gls{z}, \gls{L1}} = \sqrt{\gls{eps}_{\gls{L1}} \gls{nu}^2 - \gls{q}^2}\) (where \(\gls{eps}_{\gls{L1}}\) is the layer permittivity, \gls{nu} is the vacuum wavenumber of the light, and \gls{q} is the in-plane momentum).

Following the approach of Zhan \textit{et al.} (2013) \autocite{Zhan2013} we define a transmission matrix, \gls{T uv}, which relates the fields immediately either side of a single interface between layers \gls{L1} and \gls{L2} as:
\begin{equation}
    \begin{aligned}
        \gls{T uv}
        \begin{pmatrix}
            \gls{fwd}_{\gls{L1}} \\
            \gls{bwd}_{\gls{L1}}
        \end{pmatrix}
        \equiv                      &
        \begin{pmatrix}
            1 + {\rho}_{\gls{L1}\gls{L2}} & 1 - {\rho}_{\gls{L1}\gls{L2}} \\
            1 - {\rho}_{\gls{L1}\gls{L2}} & 1 + {\rho}_{\gls{L1}\gls{L2}}
        \end{pmatrix}
        \begin{pmatrix}
            \gls{fwd}_{\gls{L1}} \\
            \gls{bwd}_{\gls{L1}}
        \end{pmatrix}
        =
        \begin{pmatrix}
            \gls{fwd}_{\gls{L2}} \\
            \gls{bwd}_{\gls{L2}}
        \end{pmatrix}          \\
        {\rho}_{\gls{L1}\gls{L2}} = &
        \begin{cases}
            \frac{\gls{eps}_{\gls{L1}} \gls{nu}_{\gls{z}, \gls{L2}}}{\gls{eps}_{\gls{L2}} \gls{nu}_{\gls{z}, \gls{L1}}}, & \text{for p-polarization} \\
            \frac{\gls{nu}_{\gls{z}, \gls{L2}}}{\gls{nu}_{\gls{z}, \gls{L1}}},                                           & \text{for s-polarization}
        \end{cases}
    \end{aligned}
    \label{eq:trans}
\end{equation}

We also define a propagation matrix, \gls{P u}, which relates the fields at the start of layer \gls{L1}, with thickness \(\gls{d}_{\gls{L1}}\), to the fields at the end:
\begin{equation}
    \gls{P u} \equiv
    \begin{pmatrix}
        \gls{exp}^{-\gls{i} \gls{nu}_{\gls{z}, \gls{L1}} \gls{d}_{\gls{L1}}} & 0                                                                   \\
        0                                                                    & \gls{exp}^{\gls{i} \gls{nu}_{\gls{z}, \gls{L1}} \gls{d}_{\gls{L1}}}
    \end{pmatrix}
    \label{eq:prop}
\end{equation}

Then for a sample formed of layers \(0, 1, ..., N-1, N\), we can define a transfer matrix for the whole stack, via repeated matrix multiplication, as:
\begin{equation}
    M \equiv T_{01} P_{1} ... P_{N-1} T_{(N-1)N}
    \label{eq:full}
\end{equation}

Both s- and p-polarized reflection coefficients can then be obtained by
\begin{equation}
    r = \frac{M[1, 0]}{M[0, 0]}
    \label{eq:refl_coef}
\end{equation}

\subsubsection{Quasistatic}

In the quasistatic limit \autocite{Skaar2019}, the \gls{z}-component of wavevector becomes:
\begin{equation}
    \gls{nu}_{\gls{z}} = \gls{i} \gls{q}
    \label{eq:qs_lim}
\end{equation}

Thanks to momentum conservation, \gls{q} must be the same throughout the sample, meaning \(\gls{nu}_{\gls{z}}\) is also constant and no longer depends on the permittivity of a particular layer.
We can therefore replace all \(\gls{nu}_{\gls{z}, \gls{L1}}\) and \(\gls{nu}_{\gls{z}, \gls{L2}}\) with \(\gls{i} \gls{q}\) to adapt the \gls{tmm} to work for quasistatic reflection coefficients.

The transmission and propagation matrices then become:
\begin{equation}
    \begin{aligned}
        \gls{T uv}^{\mathrm{(QS)}}
        \equiv &
        \begin{pmatrix}
            1 + \frac{\gls{eps}_{\gls{L1}}}{\gls{eps}_{\gls{L2}}} & 1 - \frac{\gls{eps}_{\gls{L1}}}{\gls{eps}_{\gls{L2}}} \\
            1 - \frac{\gls{eps}_{\gls{L1}}}{\gls{eps}_{\gls{L2}}} & 1 + \frac{\gls{eps}_{\gls{L1}}}{\gls{eps}_{\gls{L2}}}
        \end{pmatrix} \\
        \gls{P u}^{\mathrm{(QS)}}
        \equiv &
        \begin{pmatrix}
            e^{q \gls{d}_{\gls{L1}}} & 0                         \\
            0                        & e^{-q \gls{d}_{\gls{L1}}}
        \end{pmatrix},
    \end{aligned}
    \label{eq:qs_matrices}
\end{equation}

Here a superscript (QS) refers to the quasistatic equivalent of an electrodynamic parameter.
These can then be combined as for the electrodynamic case above, and the quasistatic reflection coefficient can be found from:
\begin{equation}
    \gls{beta multi} = \frac{M^{\mathrm{(QS)}}[1, 0]}{M^{\mathrm{(QS)}}[0, 0]}
    \label{eq:refl_coef_qs}
\end{equation}

Note that the complex oscillation terms in the propagation matrix are replaced with exponentially decaying and growing\footnotemark{} terms (representing evanescent, non-propagating fields).

\footnotetext{
The \(e^{q \gls{d}_{\gls{L1}}}\) term can be problematic as it can lead to numerical overflow for large values of \(q \gls{d}_{\gls{L1}}\).
However we find that \(\gls{beta}(\gls{q})\) typically converges for \gls{q} values much smaller than this limit, so we can avoid this problem by simply clipping the output of \(e^{q \gls{d}_{\gls{L1}}}\) to the maximum value for double-precision floating-point numbers without noticeably effecting the results.
}

\subsection{Momentum integral discretization}
\label{sec:integrals}

In snompy, the computational efficiency of the multilayer \gls{fdm} is optimized by careful treatment of semi-definite integrals with respect to in-plane momentum, \gls{q}.
See section \ref{sec:fdm} for a full description of these parameters.

For the Hauer multilayer method, these integrals are:
\begin{equation}
    \begin{aligned}
        \left.\gls{potential}\right|_{\gls{z}=0} & = \int_0^{\infty} \gls{beta multi}(\gls{q}) e^{-2 \gls{z Q} \gls{q}} d\gls{q}, \quad \mathrm{and} \\
        \left.\gls{E z}\right|_{\gls{z}=0}       & = \int_0^{\infty} \gls{beta multi}(\gls{q}) \gls{q} e^{-2 \gls{z Q} \gls{q}} d\gls{q}
    \end{aligned}
    \label{eq:phi_E_restatement}
\end{equation}

For the charge average method this is:
\begin{equation}
    \gls{beta Q ave} = \frac{\int_0^{\infty} \gls{beta multi}(\gls{q}) \gls{q} e^{-2 \gls{z Q} \gls{q}} d\gls{q}}
    {\int_0^{\infty} \gls{q} e^{-2 \gls{z Q} \gls{q}} d\gls{q}}
    \label{eq:beta_mester_restatement}
\end{equation}

These are first rearranged and simplified, using the substitution \(\gls{X} = 2 \gls{z Q} \gls{q}\), to give:
\begin{equation}
    \begin{aligned}
        \left.\gls{potential}\right|_{\gls{z}=0} & = \frac{1}{2 \gls{z Q}}   & \int_0^{\infty} & \gls{beta multi}\left(\frac{\gls{X}}{2 \gls{z Q}}\right) e^{-\gls{X}} d\gls{X},                            \\
        \left.\gls{E z}\right|_{\gls{z}=0}       & = \frac{1}{4 \gls{z Q}^2} & \int_0^{\infty} & \gls{beta multi}\left(\frac{\gls{X}}{2 \gls{z Q}}\right) \gls{X} e^{-\gls{X}} d\gls{X}, \quad \mathrm{and} \\
        \gls{beta Q ave}                         & =                         & \int_0^{\infty} & \gls{beta multi}\left(\frac{\gls{X}}{2 \gls{z Q}}\right) \gls{X} e^{-\gls{X}} d\gls{X}
    \end{aligned}
    \label{eq:substitutions}
\end{equation}

We then make use of the Gauss-Laguerre approximation \autocite{Ehrich2002}:
\begin{equation}
    \int_0^{\infty} f(\gls{X}) e^{-\gls{X}} d\gls{X} \approx \sum_{\gls{m}=1}^M w_{\gls{m}} f(\gls{X}_{\gls{m}})
    \label{eq:gauss_lag}
\end{equation}

Here \(\gls{X}_{\gls{m}}\) is the \(\gls{m}^{\mathrm{th}}\) root of the order-\(M\) Laguerre polynomial:
\begin{equation}
    L_M(\gls{X}) = \sum_{\gls{j}=0}^M \binom{M}{\gls{j}} \frac{(-1)^{\gls{j}}}{\gls{j}!} \gls{X}^{\gls{j}}
    \label{eq:lag}
\end{equation}
and \(w_{\gls{m}}\) is a weight given by:
\begin{equation}
    w_{\gls{m}} = \frac{\gls{X}_{\gls{m}}}{\left((M + 1) L_{M+1}(\gls{X}_{\gls{m}})\right)^2}
    \label{eq:lag_weight}
\end{equation}

The integrals can therefore be approximated by the sums:
\begin{equation}
    \begin{aligned}
        \left.\gls{potential}\right|_{\gls{z}=0} & \approx \frac{1}{2 \gls{z Q}}   & \sum_{\gls{m}=1}^M & \gls{beta multi}\left(\frac{\gls{X}_{\gls{m}}}{2 \gls{z Q}}\right) w_{\gls{m}},                                      \\
        \left.\gls{E z}\right|_{\gls{z}=0}       & \approx \frac{1}{4 \gls{z Q}^2} & \sum_{\gls{m}=1}^M & \gls{beta multi}\left(\frac{\gls{X}_{\gls{m}}}{2 \gls{z Q}}\right) \gls{X}_{\gls{m}} w_{\gls{m}}, \quad \mathrm{and} \\
        \gls{beta Q ave}                         & \approx                         & \sum_{\gls{m}=1}^M & \gls{beta multi}\left(\frac{\gls{X}_{\gls{m}}}{2 \gls{z Q}}\right) \gls{X}_{\gls{m}} w_{\gls{m}}
    \end{aligned}
    \label{eq:sums}
\end{equation}

The choice of \(M\) will affect the accuracy of the approximation, with higher values requiring more memory but leading to a more accurate evaluation.
The weights and roots for a particular order, \(M\), can be precomputed to speed up execution time.

\subsection{Dielectric permittivity models}
\label{sec:eps}

\subsubsection{Hypothetical semiconductors}
\label{sec:eps_semi}

The hypothetical semiconducting samples shown in figure \ref{fig:greatest_hits}\textlet{b}, are represented as bulk samples with a Drude function permittivity, \gls{drude}, defined by:

\begin{equation}
    \gls{drude}(\gls{nu})
    = \gls{eps inf}
    - \frac{\gls{nu plasma}^2}{\gls{nu}^2 + \gls{i} \gls{gamma} \gls{nu}}
    \label{eq:drude}
\end{equation}

Here the parameter \gls{eps inf} represents the high-frequency permittivity, \gls{nu plasma} is the plasma wavenumber, and \gls{gamma} is the damping rate

The specific parameters used to model the semiconductors are given below in table \ref{tab:eps_semi}.

\begin{table}[H]
    \centering
    \begin{tabularx}{\columnwidth}{ >{\centering\arraybackslash}X  >{\centering\arraybackslash}X }
        parameter       & value                           \\
        \hline
        \gls{eps inf}   & \num{2}                         \\
        \gls{nu plasma} & \SIrange{10}{10}{\centi m^{-1}} \\
        \gls{gamma}     & \(\frac{\gls{nu plasma}}{3}\)   \\
    \end{tabularx}
    \caption{Hypothetical semiconductor permittivity parameters.}
    \label{tab:eps_semi}
\end{table}

\subsubsection{\Glsentrylong{pmma} (\glsentrytext{pmma})}
\label{sec:eps_pmma}

Figure \ref{fig:spectra}\textlet{c} shows the simulated dielectric permittivity, \(\gls{eps}_{\textrm{\gls{pmma}}}\), used throughout this paper for \gls{pmma} close to its \ce{C=O} bond.
We chose the parameters of this model to produce the best agreement with the spectra shown in figures \ref{fig:spectra}\textlet{a} and \textlet{b}, while using the \gls{fdm} parameters described by Mester \textit{et al.} in the original reference from which the spectra were taken \autocite{Mester2020}.

We represent the permittivity by an inhomogeneously broadened oscillator, defined by convolution of a Lorentzian oscillator, \gls{lorentz}, with a Gaussian kernel, \gls{gauss}, as:

\begin{equation}
    \gls{eps}_{\textrm{\gls{pmma}}}
    = \gls{gauss}(\gls{nu}) * \gls{lorentz}(\gls{nu})
    \label{eq:broadened}
\end{equation}

The Lorentzian oscillator is defined by:

\begin{equation}
    \gls{lorentz}(\gls{nu})
    = \gls{eps inf}
    + \frac{\gls{A osc}}{\gls{nu osc}^2 - \gls{nu}^2 - \gls{i} \gls{gamma osc} \gls{nu}}
    \label{eq:lorentz}
\end{equation}

Here the parameter \gls{eps inf} represents the high-frequency permittivity, \gls{A osc} is the oscillator amplitude, \gls{nu osc} is the oscillator wavenumber, and \gls{gamma osc} is the oscillator width.
The Gaussian kernel is defined as:

\begin{equation}
    \gls{gauss}(\gls{nu})
    = \sqrt{\frac{\gls{std}}{\pi}} \gls{exp}^{-\gls{std} \gls{nu}^2}
    \label{eq:gauss}
\end{equation}

Here \gls{std} is the width of the kernel, which defines degree of inhomogeneous broadening.

The specific parameters used to model \gls{pmma} are given below in table \ref{tab:eps_pmma}.

\begin{table}[H]
    \centering
    \begin{tabularx}{\columnwidth}{ >{\centering\arraybackslash}X  >{\centering\arraybackslash}X }
        parameter       & value                     \\
        \hline
        \gls{eps inf}   & \num{2.8}                 \\
        \gls{A osc}     & \SI{4.6e4}{\centi m^{-2}} \\
        \gls{nu osc}    & \SI{1730}{\centi m^{-1}}  \\
        \gls{gamma osc} & \SI{2.1}{\centi m^{-1}}   \\
        \gls{std}       & \SI{10.8}{\centi m^{-1}}  \\
    \end{tabularx}
    \caption{\Gls{pmma} permittivity parameters.}
    \label{tab:eps_pmma}
\end{table}

\subsubsection{Graphene}
\label{sec:eps_graphene}

To simulate the dielectric permittivity of graphene, \(\gls{eps}_{gr}\), we first model the frequency-, \gls{omega}-, dependent optical conductivity using a simple Drude-type response\footnotemark as:

\begin{equation}
    \gls{cond}
    = \frac{\gls{e}^2 \gls{E f}}{\pi \gls{hbar}^2}
    \left(\frac{\gls{i}}{\gls{omega} + \frac{\gls{i}}{\gls{tau}}}\right)
    \label{eq:graphene}
\end{equation}

Here \gls{e} is the electron charge, \gls{E f} is the Fermi energy of the graphene, \gls{hbar} is the reduced Planck constant, and \gls{tau} is the scattering time for graphene charge carriers \autocite{Koppens2011}.

\footnotetext{There are more accurate models of graphene's optical conductivity, such as the local and non-local random phase approximation \autocite{Koppens2011}, but the Drude response is sufficient for the simple demonstration here.}

We then convert this to a finite-thickness permittivity using:
\begin{equation}
    \gls{eps}_{gr}
    = \gls{eps inf} + \frac{\gls{i} \gls{cond}}{\gls{eps 0} \gls{omega} \gls{d}_{gr}}
    \label{eq:graphene_eps}
\end{equation}

Here the parameter \gls{eps inf} represents the high-frequency permittivity, \gls{eps 0} is the vacuum permittivity, and \(\gls{d}_{gr}\) is the thickness of graphene.

The specific parameters used to model graphene are given below in table \ref{tab:eps_graphene}.

\begin{table}[H]
    \centering
    \begin{tabularx}{\columnwidth}{ >{\centering\arraybackslash}X  >{\centering\arraybackslash}X }
        parameter        & value               \\
        \hline
        \gls{eps inf}    & \num{1}             \\
        \gls{E f}        & \SI{100}{\milli eV} \\
        \gls{tau}        & \SI{500}{\femto s}  \\
        \(\gls{d}_{gr}\) & \SI{0.3}{\nano m}   \\
    \end{tabularx}
    \caption{Graphene permittivity parameters.}
    \label{tab:eps_graphene}
\end{table}

\printbibliography

\section*{Acknowledgements}

We thank Iris Niehues, Martin Schnell and Edoardo Vicentini from CIC nanoGUNE, as well as Alexander Govyadinov from Attocube Systems AG, for their helpful comments during the preparation of snompy and this manuscript.

We would like to thank UKRI and EPSRC for research funding via projects MR/T022140/1, EP/S037438/1, and the NAME programme grant (EP/V001914/1), as well as the Henry Royce Institute for access to the CUSTOM facility at the University of Manchester (EP/T01914X/1).
We also acknowledge the support of the UK government Department for Science, Innovation and Technology through the UK National Quantum Technologies Programme.

\clearpage

\printglossary
\printglossary[type=\acronymtype]

\end{document}